\documentclass[aps,prl,twocolumn,showpacs]{revtex4}

\usepackage{epsfig}
\usepackage{graphicx}

\DeclareGraphicsExtensions{.eps,.EPS}

\begin{document}

\title{Control of dipolar relaxation in external fields}
\author{B. Pasquiou, G. Bismut, Q. Beaufils, A. Crubellier$^{\diamond}$, E. Mar\'echal, P. Pedri, L. Vernac, O. Gorceix and B. Laburthe-Tolra} %
\affiliation{Laboratoire de Physique des Lasers, CNRS UMR 7538, Universit\'e Paris 13,
99 Avenue J.-B. Cl\'ement, 93430 Villetaneuse, France}
\affiliation{$^{\diamond }$ Laboratoire Aim\'e Cotton, CNRS II, B\^atiment 505, Campus d'Orsay, 91405 Orsay Cedex, France }

\begin{abstract}
We study dipolar relaxation in both ultra-cold thermal and Bose-condensed chromium atom gases. We show three different ways to control dipolar relaxation, making use of either a static magnetic field, an oscillatory magnetic field, or an optical lattice to reduce the dimensionality of the gas from 3D to 2D. Although dipolar relaxation generally increases as a function of a static magnetic field intensity, we find a range of non-zero magnetic field intensities where dipolar relaxation is strongly reduced. We use this resonant reduction to accurately determine the $S=6$ scattering length of chromium atoms: $a_6 = 103 \pm 4 a_0$. We compare this new measurement to another new determination of $a_6$, which we perform by analysing the precise spectroscopy of a Feshbach resonance in $d$-wave collisions, yielding $a_6 = 102.5 \pm 0.4 a_0$. These two measurements provide by far the most precise determination of $a_6$ to date. We then show that, although dipolar interactions are long-range interactions, dipolar relaxation only involves the incoming partial wave $l=0$ for large enough magnetic field intensities, which has interesting consequences on the stability of dipolar Fermi gases. We then study ultra-cold chromium gases in a 1D optical lattice resulting in a collection of independent 2D gases. We show that dipolar relaxation is modified when the atoms collide in reduced dimensionality at low magnetic field intensities, and that the corresponding dipolar relaxation rate parameter is reduced by a factor up to 7 compared to the 3D case. Finally, we study dipolar relaxation in presence of radio-frequency (rf) oscillating magnetic fields, and we show that both the output channel energy and the  transition  amplitude can be controlled by means of rf frequency and Rabi frequency.
\end{abstract}


\date{\today}

\maketitle

Up to recently, only short-range isotropic interactions have played a significant role in the physics of ground state cold atoms. In contrast, dipole-dipole interactions are long-range, and anisotropic. Strong dipole-dipole interactions have profound consequences on the properties of either ultra-cold thermal or quantum degenerate Bose or Fermi gases.  Dipole-dipole interactions greatly modify the stability properties \cite{collapse} and the excitation properties \cite{excitation} of Bose condensed gases, and can even lead to new quantum phases \cite{goral}, or novel types of superfluidity \cite{baranov}. One may also use the long range character of dipolar interactions to entangle two \cite{gaetan} or more atoms, with interesting perspectives for quantum computation \cite{jaksch}.

Strong dipole-dipole interactions arise when an atomic or molecular species carries a strong permanent magnetic or electric dipole moment. Good candidates therefore include heteronuclear molecules with large electric dipole moments (which were recently produced at large phase-space densities \cite{ni}), or atoms with large electronic spin (so far erbium \cite{jabez}, dysprosium \cite{dy} and chromium). Up to now, chromium is the only species with large dipole moment for which a quantum degenerate gas has been produced \cite{Griesmaier,beaufilsBEC}. Smaller dipolar effects were also observed in a BEC of potassium for which the scattering length can be precisely tuned to zero by means of a Feshbach resonance \cite{fattoriflorence}.

Interesting new physics comes at play when one also considers the spin degree of freedom. Spin dynamics of optically trapped multi-component Bose-Einstein Condensates (also known as spinor condensates) \cite{Ho} has been observed \cite{spindynamics}. Coherent oscillations between the spin components is driven by centrally symmetric short range exchange interactions, and the total magnetization in the system is conserved. In \cite{dsk} a first dipolar effect was observed on the spin texture of a Rb spinor condensate. Dipole-dipole interaction will introduce additional new features in spin dynamics as it couples the spin degree of freedom to orbital momentum, due to its asymmetry. For example, when the magnetic dipole mean-field created by the atoms locally dominates the applied static field, spin precession occurs and the initial magnetization of the system is transferred into mechanical rotation (the equivalent of the Einstein-de-Haas effect), with appearance of structures similar to vortices \cite{edh}. Spontaneous circulation was also predicted for some experimental parameters of ground-state spinor dipolar Bose-Einstein Condensates \cite{ueda1}.

Coupling between spin and angular momenta is provided by dipolar relaxation, a two-body process in which the total magnetization changes. The main topic of this paper is to discuss how to control both the dipolar relaxation rate in cold dipolar gases and the energy released in a single dipolar relaxation event. Controlling (and possibly reducing) the rate of dipolar relaxation is interesting, as this inelastic process generally limits the lifetime of dipolar gases when all particles are not in the lowest energy state, and in particular for spinor condensates studies. (Similar limitations occur for ultra-cold Rydberg atoms \cite{rydbergplasma}). In addition, the energy released in a dipolar relaxation  event is set by the Larmor frequency of the atoms, typically much larger than the chemical potential. To observe coherent spin dynamics due to dipolar relaxation in spinor condensates,  the Larmor frequency should be of the order of the chemical potential or less. As it is technically difficult to control weak magnetic fields at the required precision, it is appealing to device new tools to control the energy in the output channel of dipolar relaxation.

In the first part of this paper, we uncover a range of magnetic field strengths for which the dipolar relaxation rate parameter of $S=3$ Cr atoms in the stretched high field seeking state $m_S=3$ is strongly reduced, both for a thermal gas and for a BEC. This reduction stems from the last node of the $s-$wave collision $S=6$ radial wave-function. We analyse the resonant reduction of dipolar relaxation as a function of magnetic field, which leads to a new determination of the $S=6$ scattering length of Cr $a_6$. Due to the proximity of a shape resonance in $S=4, l=2$, the value of the $S=4$ scattering length $a_4$ also has a significant contribution to the dipolar relaxation rate, and our analysis therefore also leads to a new estimate of $a_4$.

Then, we demonstrate that dipolar relaxation in both a BEC and a thermal gas is only due to $s$-wave collisions, provided the magnetic field is large enough. This is in contrast to elastic dipole-dipole interactions, to which all partial waves contribute due to the long range character of dipolar interactions.

We also describe experiments at low magnetic fields, where we use optical lattices to strongly confine the motion in one direction of space. In the 2D cold gases that we thus produce, the density of states at the energy of the output channel is strongly reduced compared to the 3D case, which thereby leads to a strong reduction of dipolar relaxation in the $m_S=3$ state. In addition, dipolar relaxation in the 1D lattice depends on the orientation of the magnetic field relative to the plane of the 2D gases, a direct consequence of the interplay between the anisotropy of the trapping potential and the anisotropy of dipole-dipole interactions.

Finally we describe experiments where dipolar relaxation of atoms in a BEC of $m_S=-3$ Cr atoms (usually energetically forbidden) is triggered by an rf photon. We show that the rf-assisted loss mechanism is due to dipolar relaxation between rf manifolds, and that the loss parameter and the energy released in a dipolar relaxation event can be controlled by means of the rf Rabi frequency $\Omega$ and the rf frequency $\omega$.

\section{The effect of molecular potentials on dipolar relaxation}

Dipolar relaxation has already been experimentally studied in Li \cite{lithium}, metastable He \cite{helium}, and Cr
\cite{hensler} gases. Dipolar relaxation cross sections have also been
theoretically estimated by means of coupled channels methods
\cite{Lagendijk}, or using first order perturbation theory
\cite{heliumth}. In chromium, dipolar relaxation has been
estimated using the bare first order Born approximation  ignoring all effects of the
short-range molecular potentials, and compared to experimental results
in \cite{hensler}. Here, we will show that while this approach is valid
at relatively low magnetic fields, it is crucial to take into account the
molecular potentials at higher magnetic fields.

In this part, we study the specific example of two $\left|m_S=3\right\rangle$ colliding chromium
atoms, in a magnetic field of intensity $B$, set along the $z$ direction. $S=3$ is the spin of
ground state chromium atoms. Dipolar interactions between two particles separated by $\vec r$ can be written as:
\begin{equation}
V_{dd}(\vec r)=\frac{d^2}{r^5}\left[r^2\vec S_1\cdot\vec S_2-3 (\vec S_1 \cdot \vec r) (\vec S_2\cdot \vec r)  \right]
\label{vdd}
\end{equation}
where $d^2 =\mu_0 \left(g_S \mu_B\right)^2 / 4 \pi$ and $r=\left\|\vec r\right\|$. $\mu_0$ is the magnetic constant, $g_S \approx 2$ is the Land\'e factor for ground state chromium atoms and $\mu_B$ is the Bohr magneton. Starting with a pair of atoms in state $\left|m_S=3,m_S=3\right\rangle \equiv
\left|0\right\rangle$, there are two relaxation channels due to the dipole interaction operator $V_{dd}$, channel 1:
\begin{eqnarray}
\left|3,3\right\rangle \longrightarrow \frac{1}{\sqrt{2}}
\left(\left|3,2\right\rangle+\left|2,3\right\rangle\right) \equiv
\left|1\right\rangle \label{channel1} \\
\Delta E^{(1)} = g_S \mu_B B 
\end{eqnarray}
and channel 2:
\begin{eqnarray}
\left|3,3\right\rangle \longrightarrow \left|2,2\right\rangle \equiv
\left|2\right\rangle \label{channel2} \\
\Delta E^{(2)} = 2 g_S \mu_B B  
\end{eqnarray}
$\Delta E^{(j=(1,2))}$ are the gains in kinetic energy by the pair of
atoms after a dipolar relaxation event for both channels. For atoms in the
state $m_S=3$, dipolar relaxation is exo-energetic: the kinetic energy increase is directly proportional to the
Larmor frequency.  In contrast, dipolar relaxation of chromium atoms in
the lowest state of energy $m_S=-3$ is endoenergetic, which has been used
to cool dipolar atoms \cite{fattori}.

If we characterize the relative motion of the particles by the relative wavefunctions $\Psi_{in}(\vec{r})$ and $\Psi_{out}(\vec{r})$ for the incoming state and the ouput state respectively, the two channels are described by the following matrix elements:

\begin{eqnarray}
V_1  = 3 S^{3/2} d^2 \left\langle \Psi_{out}\right|
\frac{(x+iy) z}{r^2} \frac{1}{r^3} \left| \Psi_{in} \right\rangle  \label{matrixc1} \\ \nonumber
V_2=  \frac{3}{2} S d^2 \left\langle \Psi_{out}\right|
\frac{(x+iy)^2}{r^2} \frac{1}{r^3} \left| \Psi_{in} \right\rangle \label{matrixc2} \\ 
\label{matrixelement1and2}
\end{eqnarray}
$x,y,z$ are the relative coordinates of
the two atoms separated by $\vec{r}$. $V_{j=(1,2)}$ is the matrix
element for dipolar relaxation in channel 1 or 2. As $V_1$ and $V_2$ involve terms
in $(x+iy)$ dipolar relaxation induces transitions between the different partial waves.

To estimate the cross-sections for dipolar collisions in the framework
of the first order Born approximation, one calculates the Fourier transform of the
dipole-dipole coupling independently for each channel,
$\widetilde{V_{(j)}}\left(\vec{k_f}-\vec{k_i}\right)$, for any value of
the initial and final wave-vectors $\vec{k_i}$ and $\vec{k_f}$. Then,
the total cross-section is obtained by summing
$\left|\widetilde{V_{(j)}}\left(\vec{k_f}-\vec{k_i}\right)\right|^2$
over all possible values of $\vec{k_f}$ and averaging over all possible $\vec{k_i}$.

The results of the calculation for the elastic and inelastic
cross-sections (for channel 1 and 2) are respectively \cite{hensler}:
\begin{eqnarray}
\sigma_0&=&\frac{16\pi}{45}S^4\left(\frac{d^2 m}{\hbar^2}\right)^2 f(1)
\label{sigma0born} \\
\sigma_1&=&\frac{8\pi}{15}S^3\left(\frac{d^2 m}{\hbar^2}\right)^2
f(k_f^{(1)}/k_i)k_f^{(1)}/k_i \label{sigma1born} \\
\sigma_2&=&\frac{8\pi}{15}S^2\left(\frac{d^2 m}{\hbar^2}\right)^2
f(k_f^{(2)}/k_i)k_f^{(2)}/k_i \label{sigma2born} \\ \nonumber
\end{eqnarray}
where
\begin{equation}
f(u)=1 + \epsilon
\left(-\frac{1}{2}-\frac{3}{8}\frac{(1-u^2)^2}{u(1+u^2)}\log\left(\frac{(1-u)^2}{(1+u)^2}\right)\right)
\end{equation}
$\epsilon =1$ for bosons (as in \cite{hensler}), and $\epsilon =-1$ for
fermions. $k_i$ and $k_f^{(j)}$ are related through the conservation of
energy equation:
\begin{equation}
\frac{\hbar^2 (k_f^{(j)})^2}{m}=\frac{\hbar^2 k_i^2}{m} + \Delta E ^{(j)}
\end{equation}
where $m$ is the atom mass.

Eq. (\ref{sigma0born}) describes the elastic cross section due to dipole
dipole interactions. It is worth emphasizing that this cross section
does not vanish at small collision energy for fermions, showing that non
zero partial waves contribute to the elastic cross section even at zero
collision energy (as polarized fermions do not interact in $s-$wave), a
direct consequence of the long range character of dipole-dipole
interactions. In addition, it is enlightening to emphasize that
$\sqrt{\sigma_0}$ is, to within numerical factors, equal to the range of
the dipole-dipole interaction potential $R_{dd}$, defined by
$\frac{\hbar^2}{m R_{dd}^2}= \frac{d^2 S^2}{R_{dd}^3}$.

Eq. (\ref{sigma1born}) and (\ref{sigma2born}) give respectively the
cross-sections for dipolar relaxation in channels 1 and 2.
As we shall later see, the experiment does not directly
provide a measurement of cross sections, but rather a measurement of the
corresponding loss rate parameters. For one particle moving at velocity $v_i$ in a still medium of
density $n$, the loss rate is $\Gamma_j^{1} = 2 n \sigma_j v_i = 2 n
\sigma_j \hbar \frac{k_i}{m/2}$. The factor 2 is due to the fact that
two atoms are involved: both are lost in a  single dipolar relaxation event when the gain in energy of each atom is
higher than the trap depth. To relate this rate parameter to the rate parameter in a gas of $N$
particles, one should multiply this result by $N/2$ to properly account for all pairs of particle. The final loss rate parameter is therefore
$\beta_r^j = \sigma_j v_i = 2 \sigma_j \hbar \frac{k_i}{m}$.

The main weakness of the calculation shown above is that it neglects all
other interatomic potentials (i.e. the electrostatic couplings). At low
collision energies, centrifugal barriers $\hbar^2 l(l+1) / m r^2$
for particles colliding in partial waves $l>0$ prevent the particles
from getting close  to one another, so that one can indeed safely
neglect the electrostatic potentials ($C_6/r^6$, $C_8/r^8$, ...).
However, particles colliding in $s-$wave  can approach each other and
experience the short range interaction potential. To take this into account,
it is useful to develop the collision wave function in the partial wave
basis.
\begin{figure}
\centering
\includegraphics[width= 2.8 in]{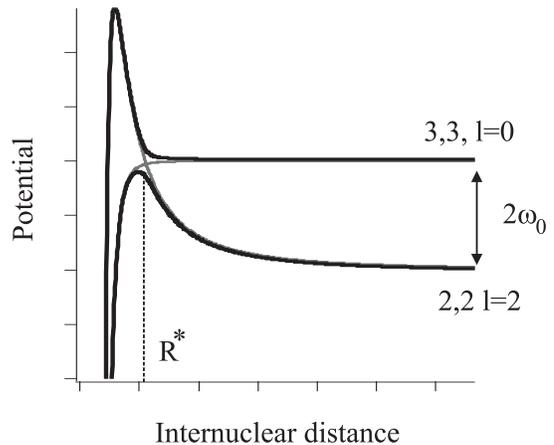}
\caption{\setlength{\baselineskip}{6pt} {\protect\scriptsize
Dipolar relaxation from the molecular physics point of view, illustrated for channel 2 (sketch); the molecular curves of the initial $l=0$ potential and the final $l=2$ potential  cross at interparticle distance $R^*$.}}
\label{principe}
\end{figure}
At long distances ($i.e.$ further
than the van der Waals range of the potential $R_{vdW}$, defined by
$\hbar^2/(m R_{vdW}^2)= C_6/R_{vdW}^6$), there are analytical
solutions to the Schr\" odinger equation describing the relative motion of
two particles, for any partial wave. In a first approach, we will use these analytical wavefunctions to calculate the dipolar relaxation rate parameter. As we shall see, this approach (although inaccurate) will provide some physical insight into the mechanism at play in dipolar relaxation.

For a spherical gaussian BEC, the situation is relatively simple, as the symmetry
of the order parameter insures that the reduced two-body wave function
has a pure $s-$wave character. In a BEC, dipolar relaxation
couples the purely $s-$wave collision wave function to $l=2$. Indeed,
the matrix element for both channels of dipolar relaxation are
proportional to a spherical harmonics $Y_2^j$ (see eq. (\ref{matrixc1})
and (\ref{matrixc2})), and thus couple $l=0$ to $l=2$ by the following
angular matrix elements:
\begin{eqnarray}
V_1^{0\leftrightarrow2}(r) & = & \left\langle 0,l=0
\right|V_{dd}\left|1,l=2,m_l=1\right\rangle \nonumber \\ & = & 3 S^{3/2} d^2
\sqrt{\frac{2}{15}} \frac{1}{r^3} \nonumber \\
V_2^{0\leftrightarrow2}(r) & = & \left\langle 0,l=0
\right|V_{dd}\left|2,l=2,m_l=2\right\rangle \nonumber \\ & = & 3 S d^2 \sqrt{\frac{2}{15}}
\frac{1}{r^3} 
\label{matrixelement}
\end{eqnarray}

Having thus determined the matrix elements for the orbital coupling from $l=0$ to $l=2$,
we now turn to the evaluation of the radial coupling of
the corresponding wavefunctions. At zero temperature, and in free space, the
incoming radial wave function at sufficiently large distances ($i.e.$ larger than $R_{vdW}$) can be written as:
\begin{equation}
F_{in}(r)=\sqrt{\frac{4 \pi}{v_0}}\left(1-\frac{a_6}{r}\right) \label{psiin}
\label{l0}
\end{equation}
where $v_0$ is a normalizing constant (see below).
At large distances, the energy-normalized $l=2$ output radial wavefunction
reads \cite{landau}:
\begin{equation}
F_{out}^{(j)}(r)= \frac{2 \sqrt{\pi m k_f^{(j)}}}{h} j_{2}(k_f^{(j)} r)
\label{psiout}
\end{equation}
$j_{2}$ is the second spherical Bessel function, and $k_f^{(j)}$ are the
final wave vectors for each channel of dipolar relaxation $\hbar^2
(k_f^{(j)})^2/m= \Delta E^{(j)}$. The coupling between $F_{in}$ and
$F_{out}$ is given by the Fermi golden rule \cite{who}:
\begin{equation}
\hbar \Gamma^{(j)} = 2 \pi \left(\int_0^{\infty} F_{in}(r)
F_{out}^{(j)}(r) V_j^{0\leftrightarrow2}(r) r^2 dr \right)^2
\label{minanalytique}
\end{equation}
$\Gamma^{(j)}$ (which can be calculated analytically) is the event
rate associated to the dipolar relaxation in channel $(j)$ \cite{who}. It is now
important to clarify the value of the normalization constant $v_0$.
This can be done for the example of an isotropic harmonic trap with
trapping frequency $\omega /2 \pi$. At long distances, the radial
wave function is essentially set by $\exp(-r^2/4a_{HO}^2)$, where
$a_{HO}=\sqrt{ \hbar/(m \omega)}$, whereas at short distances (but
not too short, as explained above), it is set by $F_{in}(r)$. As
$a_{HO}>>a_6$, one can safely approximate the initial $l=0$ radial wave function in the trap by $F_{in}^{trap}(r)=F_{in}(r)
\exp(-r^2/4a_{HO}^2)$. Using this form for the incoming wave function, one
can now calculate the coupling between $F_{in}^{trap}$ and
$F_{out}$ by an equation analogous to eq. (\ref{minanalytique}). The
integral is still analytical, and almost identical to the result of eq. (\ref{minanalytique}), for our
experimental parameters. We therefore use $F_{in}$ of eq. \ref{l0} for the
calculation of the dipolar relaxation coupling, and $F_{in}^{trap}$
for the determination of $v_0$. Normalizing
$F_{in}^{trap}$ using $\int \left|F_{in}^{trap}\right|^2 r^2 dr = 1$, we find $v_0= \left(a_{HO}\right)^3 \left( 2 \pi \right)^{3/2}$.

To now relate $\Gamma^{(j)}$, calculated for a pair of atoms, to the
value of the rate parameter for dipolar relaxation $\beta_r$, we recall the definition of $\beta_r$, through the local equation :$
\frac{dn}{dt} \equiv -\beta _r n^2$. Integrating this equation over the
 profile corresponding to the density probability of two
particles in the vibrational ground state of the trap, we find that the
associated loss rate is $2 \Gamma = \frac{n_0}{2^{3/2}} \beta_r$ ($\Gamma$
is the collision event rate so that $2 \Gamma$ is the loss rate), where $n_0 =2
\left(\frac{m \omega}{\pi \hbar}\right)^{3/2}$, so that $\Gamma =
\frac{1}{v_0} \beta_r$. As a consequence, we obtain:
\begin{eqnarray}
\beta_{r,1}^{0\leftrightarrow2} = \Gamma^{(1)} v_0  = \frac{16 \pi}{15}
\frac{m}{\hbar^3} S^3 d^4 \left(1-\frac{3 a_6 k_f^{(1)}
\pi}{16}\right)^2 k_f^{(1)} \label{beta1} \\
\beta_{r,2}^{0\leftrightarrow2} = \Gamma^{(2)} v_0  = \frac{16 \pi}{15}
\frac{m}{\hbar^3} S^2 d^4 \left(1-\frac{3 a_6 k_f^{(2)}
\pi}{16}\right)^2 k_f^{(2)}
\label{beta2}
\end{eqnarray}

In the limit where $k_i \rightarrow 0$ (BEC case) and $a_6 =0$ (no molecular potential), these equations exactly correspond
to half the value of the results of the rate parameter
associated to the cross-sections given in eqs. (\ref{sigma1born}) and (\ref{sigma2born}) (corresponding to \cite{hensler}). The difference of a factor of two arises from  the different way of symmetrizing a wavefunction when particles are in different states (thermal gas) or in the same state (BEC) (as explicited in Annexe I). It corresponds to the difference in the second order correlation function for a thermal gas and for a BEC. 

\begin{figure}
\centering
\includegraphics[width= 2.8 in]{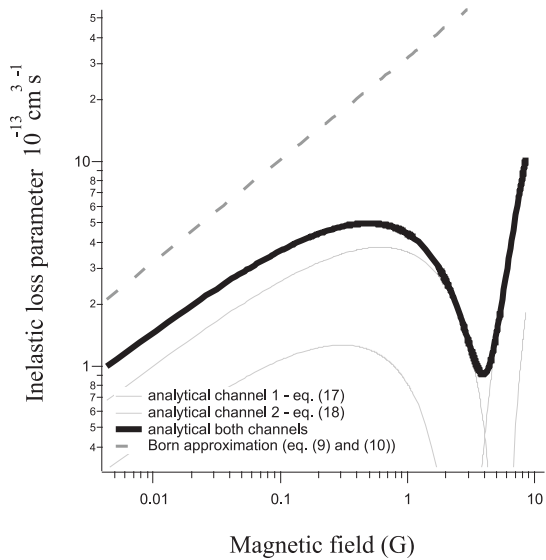}
\caption{\setlength{\baselineskip}{6pt} {\protect\scriptsize
Comparison between the analytical calculation taking into account the effect of $a_6$ (eq. \ref{beta1} and \ref{beta2}) and calculations with no molecular potentials (eq. \ref{sigma1born} and \ref{sigma2born}).}}
\label{analytiquenumerique}
\end{figure}

When $a_6 > 0$ a cancelation occurs, for each channel of dipolar relaxation,
for a given value of $k_f$, $i.e.$ for a given value of the magnetic
field. These two separate cancelations result in a dip in the
dipolar relaxation loss parameter as a function of magnetic field. This
dip is the main feature missing when calculating the cross sections for
dipolar relaxation while ignoring the effect of the molecular
potentials, and was already discussed in \cite{heliumth}, although never
observed experimentally to our knowledge. Figure \ref{analytiquenumerique} shows the results of $\beta_{r,1}^{0\leftrightarrow2}+\beta_{r,2}^{0\leftrightarrow2}$ from eq. (\ref{beta1}) and (\ref{beta2}) along with the dipolar relaxation rate parameter calculated from eq. (\ref{sigma1born}) and (\ref{sigma2born}). Figure  \ref{analytiquenumerique} illustrates both the resonant dip in dipolar relaxation, and, far from the dip, the reduction by a factor of two in BECs due to the difference between the second order correlation function of a thermal gas and of a BEC at short range. 

To derive more physical intuition from equations (\ref{beta1})
(\ref{beta2}), let us notice that $\left(1-\frac{3 \pi a_6 k_f^{(j)}}{16
}\right)$ is proportional to the value of $F_{in}(r)$ for $r=\frac{16
}{3 \pi k_f^{(j)}}$. Dipolar relaxation rates will therefore be proportional
to the probability of presence of two particles at distance

\begin{equation}
R_{DR}=\frac{16}{3 \pi k_f^{(j)}}
\label{distancerelaxation}
\end{equation}
$R_{DR}$ is close to the interparticle
distance $R^*$ at which the molecular potential of the output channel crosses
the input channel molecular potential: $R^*= \hbar \sqrt{\frac{l(l+1)}{m \Delta E^{(j)}}}$, (see Figure \ref{principe}).
Despite the long range character of dipolar interactions, dipolar relaxation appears as localized, taking place at a specific interatomic range $R_{DR} \propto 1/\sqrt{B}$.

At very small magnetic fields, when $k_f a_6 \ll 1$, our results are close to the ones given in \cite{hensler} (except for a factor of 2, as explained above). $k_f a_6 \ll 1$ corresponds to a situation where dipolar relaxation occurs at a distance $R_{DR}$ much larger than $a_6$. We therefore indeed expect the molecular potentials to have no effect on dipolar relaxation: for $l=0$ the zero energy collision wavefunction at distances $r>>a_6$ is flat and independent of $a_6$. In addition, at low energy the phase shift associated to collisions in $l=2$ vanishes.

When the magnetic field gets larger, the effect of the scattering length on the incoming wave-function starts playing a role in dipolar relaxation. In our model, the cancelation of dipolar relaxation in a given channel occurs when $R_{DR} = a_6$, $i.e.$ when dipolar relaxation occurs at an interatomic distance at which lies the last node in the incoming wave-function. However, as in the case of chromium $a_6 \sim R_{vdW}$, the analytical forms for the collision wave functions given by eq. (\ref{psiin}, \ref{psiout}) are not valid in this region. At short
internuclear distances (most important at large magnetic fields), one needs to numerically calculate the incoming and output wavefunctions. We describe in the next part of this paper our experimental data and how we perform such numerical calculations to fit the results of such experiments.

\section{Measurements of dipolar relaxation and scattering lengths}

We now turn to the experimental study of dipolar relaxation. We shall see that a comparison between our experimental data and a numerical model taking into account the effect of the molecular potentials provides new estimates of Cr scattering lengths. This method has similarities with those used to interpret photoassociation experiments, in which modulations in the intensity of the photoassociation lines reflect the structure of the colliding wavefunction \cite{Miller}, which has been used for a determination of scattering lengths \cite{Cote}. Moreover, our method probes the collision wavefunction at any interatomic distance, because dipolar relaxation can be measured at any magnetic fields, whereas photoassociation relies on a resonant condition with a molecular excited state, so that it only provides discrete information on the collision wavefunction.

To study dipolar relaxation in a $m_S=3$ BEC, we first produce Cr BECs in the $m_S=-3$ state, as described in \cite{beaufilsBEC}. BECs are produced in a crossed optical dipole trap. Because trapping is weak at the end of the evaporation stage, we also use a magnetic field gradient along the vertical direction to compensate for gravity. After the BEC is produced in the crossed optical dipole trap, we recompress the trap to a value such that magnetic levitation is not necessary anymore to compensate for gravity. We then remove the vertical magnetic gradient, and we apply a bias field in the horizontal plane. We have checked that the number of atoms in the BEC - typically 10 000 - is independent of the value of the final bias field. For this part of the paper, the value of the bias field is such that the Larmor frequency of the atoms is larger than the trap depth: dipolar relaxation results in losses.
\begin{figure}
\centering
\includegraphics[width= 2.8 in]{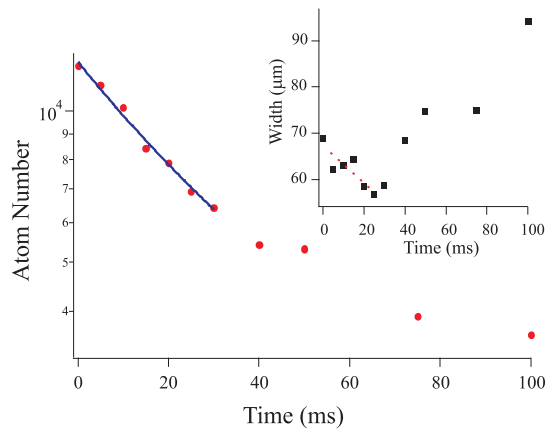}
\caption{\setlength{\baselineskip}{6pt} {\protect\scriptsize
Atom number and size of the cloud as a function of the delay time $t$ spent in $m_S=3$.}}
\label{atomlosses}
\end{figure}

We then apply a pulse of radio-frequency magnetic field linearly polarized along the vertical direction. The rf frequency is swept across the rf resonance (ie around the Larmor frequency). The rf power and the duration of the pulse (typically 2 ms) are such that more than 95 percent of the atoms end up in the $m_S=3$ state, as measured by Stern and Gerlach experiments.

After a given delay time $t$, we apply a second rf sweep to the atoms, to bring them back to the $m_S=-3$ state. We then abruptly (within 300 ns) release the optical trap, and take an absorption image after a few milliseconds (typically 5 ms) of expansion. If the delay time $t$ is short enough, we recover a BEC, with the same number of atoms and the same Thomas-Fermi radius than the initial BEC. As we do recover a BEC for delay times considerably longer than the oscillation time in the trap, our experiment is (to our knowledge) the first experiment to date to have produced a BEC of chromium atoms in the $m_S=3$ state.

As the delay time $t$ increases, the number of atoms decreases due to dipolar relaxation (see fig \ref{atomlosses}), without noticeable heating. This contrasts with what is typically observed in a trapped thermal gas: in that case, loss of atoms due to a density-dependent process mostly occurs at the center of the trap, so that the average energy lost per lost atom is less than the average energy of an atom in the sample. Because of this, inelastic losses in a trapped thermal gas results in heating of the cloud. In a BEC the chemical potential is flat over the size of the BEC, so that the energy lost per lost atom is independent of the position of this atom in the trap. Losses do not lead to heating, provided the size of the cloud adiabatically adapts to the change in number of particles, $i.e.$ provided the loss rate is small compared to the rate of oscillation in the trap. The fact that we observe losses without heating is therefore an indirect signature of the fact that the chemical potential is constant over the size of the cloud, a pre-requisite for phase coherence in the gas.

As the number of atoms decreases, the Thomas Fermi radius decreases accordingly (see the inset of fig \ref{atomlosses}). After typically 20 ms, the number of atoms gets so small that the thermal fraction increases, and we eventually completely lose the BEC: this leads to an increase of the width of the cloud after some time of flight (see the inset of fig \ref{atomlosses}). A typical set of data is shown in fig \ref{atomlosses}. We analyze such data for delay times $t$ short enough that the BEC is conserved, with no discernable thermal fraction (30 ms in the present case). The local decay equation reads:
\begin{equation} \label{decayn}
\frac{dn}{dt}=-\beta_r n^2 -\Gamma_1 n
\end{equation}
where $\beta_r$ is the loss parameter associated to dipolar relaxation, and $\Gamma_1 \approx 0.1$ s$^{-1}$ is the one-body loss coefficient due to collisions with the background gas, which is independantly measured. Typically, $\beta_r n$ is so large that $\Gamma_1$ can be neglected in eq. (\ref{decayn}). To relate this equation to the decay equation for the total number of atoms, we assume that all atoms are in the BEC, in the Thomas Fermi regime; we also assume that the scattering length in the $m_S=3$ state is identical to the scattering length in the $m_S=-3$ state. Indeed, atoms being in the stretched state for both cases, the collision is purely in the $S=6$ molecular channel, and both states therefore have the same scattering length $a_6$ provided there is no Feshbach resonance in this range of magnetic fields. Integrating eq. (\ref{decayn}) over the Thomas-Fermi density profile (and assuming that there is no thermal fraction), we find that the equation for the total number of atoms reads:
\begin{equation} \label{decayN}
\frac{dN}{dt}=- \alpha \beta _r N^{7/5} -\Gamma_1 N
\end{equation}
where $\alpha=\frac{15^{2/5}}{14 \pi}\left(\frac{m\overline{\omega}}{\hbar \sqrt{a_6}}\right)^{6/5}$ (see for example \cite{dalibard}). $\overline{\omega}$ is the geometrical mean of the oscillation frequencies at the bottom of the trap, independently measured by parametric excitation. $a_6 = 112$ $a_0$ was measured in \cite{werner}. There is an analytical solution for equation \ref{decayN}, which we use to fit our experimental data (see fig \ref{atomlosses} for example)  in order to extract a value of $\beta_r$ for each magnetic field.
\begin{figure}
\centering
\includegraphics[width= 2.8 in]{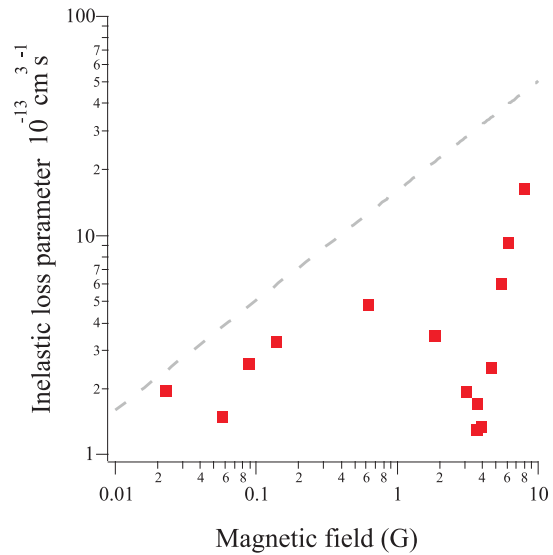}
\caption{\setlength{\baselineskip}{6pt} {\protect\scriptsize
Dipolar relaxation rate parameter as a function of magnetic field for a $m_S=3$ BEC. Squares: experimental data points. Dashed line: calculation in the first order Born approximation, with no molecular potential (eq. \ref{sigma1born} and \ref{sigma2born}); the calculated rate parameter is divided by 2 to take into account the fact that the experiment is performed with a BEC.}}
\label{antiresonance}
\end{figure}
Results of this procedure are presented fig \ref{antiresonance} as a function of the magnetic field. As predicted in the previous part of this paper, we observe a dip in the experimental dipolar relaxation rate parameter as a function of magnetic field, due to the effect of molecular potentials. As a first step to quantify this effect, we compare our experimental data to the results of the analytical model presented above. In the limit of zero collision energy, the analytical model offers direct link between $a_6$ and the magnetic field at which losses are minimum
\begin{equation} \label{miniana}
a_6 = \frac{16 (2+S) \hbar}{3 \sqrt{m g_S \mu_B B_{min}} \pi (2 \sqrt{2}+S)}
\end{equation}
We therefore may deduce from our measurement a first value of $a_6 = 117$ $a_0$. However, since this value is not much larger than $R_{vdW}$ the analytical model is not valid, as explained above. We therefore now turn to our numerical simulations, to accurately interpret the experimental data.

In our numerical simulations, the molecular potential is accounted for in the whole asymptotic region, and for both the initial and final states. The atomic spins are coupled, with total spin $\vec S_t=\vec S_1+\vec S_2$ and the dipole-dipole interaction can be written as (\cite{d-wave-Fesh})
\begin{eqnarray}
\label{Hdd-tensoriel}
H_{dd}=\frac{C_d}{r^3},
\end{eqnarray}
with
\begin{equation}
C_d=-4\sqrt 6 \mu_B^2 \frac{\mu_0}{4 \pi} \{ \vec S_1.\vec S_2 \}^{(2)}_0,
\label{angpart}
\end{equation}
where the projection of the tensor rank is taken on the internuclear axis.
Separating angular and radial parts, as allowed by the Born-Oppenheimer approximation, the relevant initial state of an atomic pair can be written as
\begin{eqnarray}
\label{cont}
|in> = |S_t,M_t,\ell,m_l>~F_{in} (r),
\end{eqnarray}
and the $j^{th}$ final molecular state after dipolar relaxation is given by
\begin{eqnarray}
\label{lie}
|out^{(j)}> = |S_t^{(j)},M_t^{(j)},\ell ^{(j)},m^{(j)}>~F_{out}^{(j)} (r),
\end{eqnarray}
with $M_t^{(j)}+m_{\ell}^{(j)}=M_t+m_l$ and $|\ell^{(j)}-\ell|=0,2$ with the exception of $\ell^{(j)}=\ell=0$. For two $m_S=3$ atoms, $S_t=6,M_t=6$, and in the case of an initial BEC one additionnally has $\ell=0,m_l=0$. 

For the corresponding final state, one has $\ell^{(j)}=2$ and there are two possible $S_t$ values, 6 and 4. The initial state is thus coupled with three final molecular states: $|out^{(1)}>=|6,5,2,1>$, with $\Delta E^{(1)}=g_S\mu_B B$, and $|out^{(2)}>=|6,4,2,2>$, $|out^{(3)}>=|4,4,2,2>$, with $\Delta E^{(2)}=2g_S\mu_B B$. State $|1>$ introduced above for our simplified analytic model corresponds to $|out^{(1)}>$, whereas $|2>$ is a linear superposition of $|out^{(2)}>$ and $|out^{(3)}>$.

The angular part of the matrix element of eq. (\ref{Hdd-tensoriel}) is evaluated using standard tensor operator technique, remembering that the angular momentum projections appearing there are taken on a fixed axis (determined by the static magnetic field). A general formula is given in  \cite{d-wave-Fesh}.

As in ref \cite{d-wave-Fesh}, the radial wavefunctions are described in a simple, purely asymptotic model, based on the concept of nodal lines \cite{crubellier,vanhaecke}. The Schr\" odinger equation for the radial collision wavefunctions is solved in the asymptotic part only by inward numerical integration starting from large $r$ values. The constraint is that the wavefunction vanishes at a given nodal position, which characterizes entirely the inner part of the potential (in the spirit of the quantum defect theory). The model depends thus on the asymptotic potential, i.e. essentially on the van der Waals $C_6$ constant (and, to a much smaller extent, on the value of $C_8$), and on the location of the chosen nodal lines. It can be shown that, in a first approximation, the position $R_0$ of the nodes depends both on the collision energy $\epsilon$ and on the rotational angular momentum $\ell$ with
\begin{equation}
R_0=R_{00}+A \epsilon +B\ell (\ell+1).
\label{nodal-lines}
\end{equation}
The constants $R_{00}$, $A$ and $B$ characterize the inner part of the different molecular potentials, and are different for each $S_t$ value. They can be deduced from experiment and we have used here values of $A$ and $B$ for $S_t=6$ and $S_t=4$  derived from the data of \cite{werner}. $A$ and $B$  act here as small correction terms. The $C_6$ and $C_8$ coefficients were also taken from \cite{werner}. The knowledge of the constant $R_{00}$ is  equivalent to the one of the scattering length of the considered molecular potential and $R_{00}$ is thus used as an adjustable parameter.

Taking into account the confinement of the particles, the initial radial wavefunction $F_{in}(r)$ behaves at large distance like a trapped atom pair wavefunction. At short distance, the interaction between the two atoms cannot be ignored and the wavefunction is proportional to a vibrational wavefunction of the $S_t=6$ molecular potential; its amplitude is fixed, through the normalization condition, by the parameters of the long-range wavefunctions. The final wavefunction $F_{out}^{(j)}(r)$, is, at long-range, an energy-normalized free wavefunction and, at short-range, a molecular  wavefunction of the $S_t^{(j)}$ potential.

The Fermi Golden rule for a given final state $(j)$
reads
\begin{eqnarray}
\Gamma^{(j)} & =  &  \frac{2 \pi}{\hbar} |<in|V_{dd}|out^{(j)} >|^2 \nonumber \\
& = & \frac{2 \pi}{\hbar} |<S_t,M_t,\ell,m_l|C_d|S_t^{(j)},M_t^{(j)},\ell^{(j)},m^{(j)}> |^2 \nonumber \\
& & I(\Delta E^{(j)})^2,
\label{fermi}
\end{eqnarray}
with
\begin{eqnarray}
I(\Delta E^{(j)})=\int_0^\infty{\frac{F_{in} (r)F_{out}^{(j)} (r)}{r^3}~ r^2dr}.
\label{rad-int}
\end{eqnarray}

The dipolar relaxation rate is the sum of the Fermi golden rule results for dipolar relaxation towards the three final states. These include the relevant different angular weigths for the three relevant final states, which are in the ratio (5.400, 0.9818, 0.8182). As a consistency check, we observe that ratio of the coupling to $|out^{(1)}>$ to the sum of the couplings to $|out^{(2)}>$ and $|out^{(3)}>$ is exactly S, equal to the ratio between dipolar relaxation in channels 1 and 2 defined in our analytical model (see eq. \ref{matrixelement}).

The variation of the integrals with $\Delta E^{(j)}$ (i.e. with magnetic field), governs dipolar relaxation. At very small values of $B$, the radial integral mostly builds at large interatomic distances. The value of the integral is determined by the behavior of the wavefunction at long distance, and the molecular potentials do not play any significant role. At higher magnetic fields values, on the other hand, the inner part of the wavefunctions contribute, and dipolar relaxation is modified by the molecular potentials, both initial and final. These general features, already revealed in the analysis of our analytic model,  still hold in our numerical model. Since this models correctly takes into account the inner part of the wavefunction (which the analytical model failed to do), we will now use it to analyze the experimental data.

\begin{figure}
\centering
\includegraphics[width= 2.8 in]{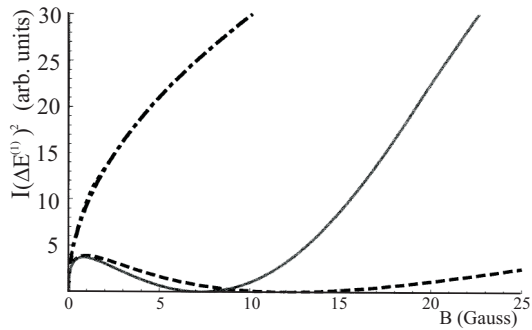}
\caption{\setlength{\baselineskip}{6pt} {\protect\scriptsize
Influence of initial and final molecular potentials on dipolar relaxation to state $|out^{(1)}>$. The figure shows the variation of the square of the radial integrals $I(\Delta E^{(1)})$ as a function of the energy gap $\Delta E^{(1)}$ (in G). The case where both potentials are ignored is the dotted-dashed line. The dashed curve corresponds to the case where the initial potential is taken into account: the strong diminution, as compared to the first curve, comes from the oscillations of the vibrational wavefunction. The final molecular potential induces an enhancement of the integral (solid line), which corresponds to an increase of the density probability close to the rotational barrier.}}
\label{molpots}
\end{figure}

In figure \ref{molpots}, we compare radial integrals describing dipolar relaxation to state $|out^{(1)}>$  calculated without molecular potentials with similar integrals calculated taking into account either the initial or both initial and final molecular potentials. (Dipolar relaxation to other final states is omitted here for better clarity.) In the first case (no molecular potentials), the initial pair wavefunction is a radial trapped pair wavefunction (see Annexe I) and the final one is an energy-normalized free pair wavefunction given by equation (\ref{psiout}). In the other cases, the same long-range wavefunctions are smoothly connected to  molecular wavefunctions, with $S_t=6$ for both the initial and final states. The $C_6$ value (733~atomic units) is taken from  \cite{werner}.

It results from figure \ref{molpots} that it is important to take into account the molecular potential both for the initial state and the final state. The fact that one needs to take into account the molecular potential in the initial state is also predicted by the analytical model presented above. However, while the analytical model predicted a zero of dipolar relaxation for approximately 5.5 G, the numerical model (ignoring the molecular potential in the final state) leads to a zero of dipolar relaxation in this channel at 12 G. This disagreement confirms that the choice of eq. (\ref{psiin}) for the incoming wavefunction is too naive, and that one needs to take into account oscillations of the wavefunction at interparticle distances smaller than $a_6$.

We also see from figure \ref{molpots} that the molecular potential cannot be ignored in the output channel either. This happens when the energy gain in dipolar relaxation is not much smaller than the height of the centrifugal $l=2$ barrier. Then, one cannot neglect the effect of the short range molecular potential, so that the choice of eq. \ref{psiout} for the output channel wavefunction is also too naive.

\begin{figure}
\centering
\includegraphics[width= 2.8 in]{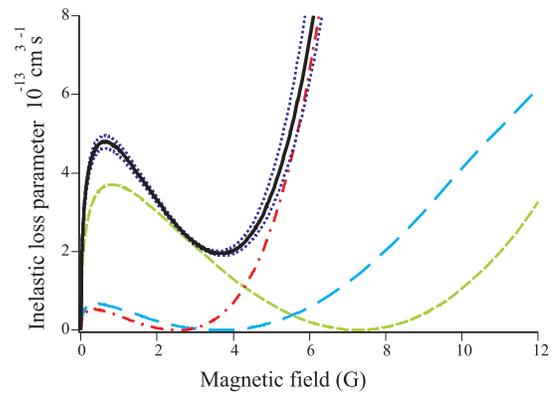}
\caption{\setlength{\baselineskip}{6pt} {\protect\scriptsize
Contribution of the three different channels of dipolar relaxation, to states $|out^{(1)}>$, $|out^{(2)}>$ and $|out^{(3)}>$, and influence of the $C_6$ coefficient. The figure shows the variation of the  relaxation rate parameter of the three final states (green dashed: $|out^{(1)}>$; blue long-dashed: to $|out^{(2)}>$; red dot-dashed: to $|out^{(3)}>$), together with their sum (thick solid line), as a function of the magnetic field. The central black solid line corresponds to the central $C_6$ value 733 $a_0$, whereas the top dotted black line corresponds to 803 $a_0$ (bottom dotted black line: 663 $a_0$).  The value of $a_4$ is set to 69~a$_0$.}}
\label{influence-C6}
\end{figure}

In figure \ref{influence-C6}, we show the contributions of the different molecular channels to dipolar relaxation, as well as their sum. The calculation shows a dip in dipolar relaxation, similar to our experimental observation. As explained above, both the initial and the final molecular potential need to be taken into account for comparison between theory and experiment.

In particular we find that the value of $a_4$ has a considerable effect on dipolar relaxation for fields larger than 1~G. We explain this rather unexpected large effect of the $S_t=4$ potential on dipolar relaxation by the presence of a shape resonance. We have performed calculations in a "universal" truncated $R^{-6}$ potential, consisting in a $R^{-6}$ potential limited at short range by an infinite repulsive wall \cite{beatriz}. In the case of chromium, we expect a $\ell=2$ shape resonance for an interval of $a_4$ 45-70~a$_0$, which incudes the 52-64~a$_0$ range recommended in \cite{werner}. In this interval, one expects a resonant increase, by tunneling through the rotational barrier, of the probability amplitude at short range and therefore of dipolar relaxation to this state. The resonance is expected close to the top of the barrier and the enhancement occurs at relatively relatively large fields. 

\begin{figure}
\centering
\includegraphics[width= 3 in]{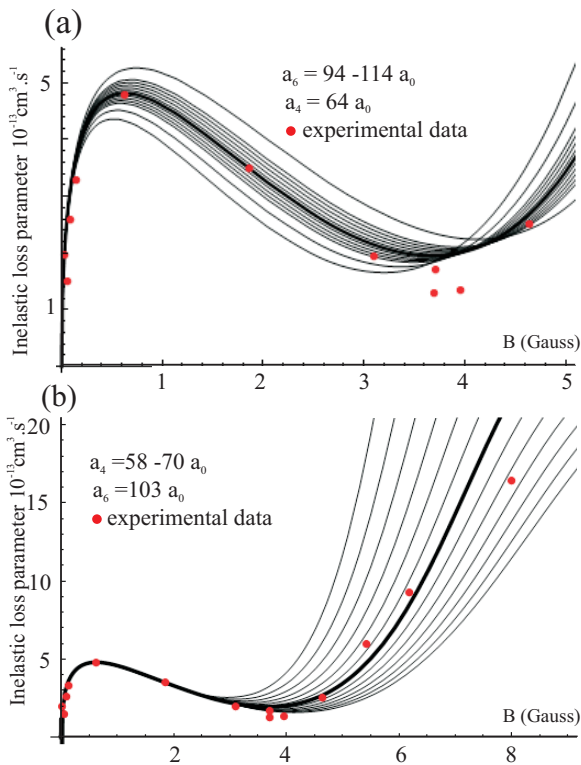}
\caption{\setlength{\baselineskip}{6pt} {\protect\scriptsize
Influence of the $a_6$ and $a_4$ values and comparison with experiment. The figure shows the variation of the dipolar relaxation rate parameter as a function of the magnetic field (in Gauss). Bullets: experimental data points. (a) Straight lines: result of the calculation for various values of $a_6$: 94 $a_0$, 98 to 107 $a_0$ with step 1~a$_0$, 110 and 114 $a_0$. The $C_6$ coefficient is here equal to 733~a$_0$, and $a_4 = 64 a_0$. (b) result of the calculation for various values of $a_4$ from 58 $a_0$ to 70 $a_0$ with step 1~a$_0$. The $C_6$ coefficient is here equal to 733~a$_0$, and $a_6 = 103 a_0$. }}
\label{a6a4}
\end{figure}

Hence, fortunately, as shown in Fig \ref{a6a4}, varying $a_4$ induces changes in dipolar relaxation only on the large magnetic field side of the dip. We therefore vary the $a_6$ value in the range $99 ~a_0$ to $107 ~a_0$, and look for the best agreement between theory and experiment for the low magnetic field side of the dip, where $a_4$ has little effect on dipolar relaxation. From this first analysis, we can deduce a new value for the $a_6$ coefficient: $a_6= (103 \pm 4) ~ a_0$. Setting $a_6$ to 103 $a_0$, we can now vary $a_4$ to find  the best agreement between theory and experiment on the 'blue side' of the dip. We thus also deduce from our analysis a new determination of $a_4 = (64 \pm 4) ~ a_0$. We also varied the $C_6$ coefficient in the range recommended in \cite{werner}, as shown in Fig \ref{influence-C6}, and we found that its effect on the reduction in dipolar relaxation is weaker than the effects of $a_4$ and $a_6$. The error bars that we give in our new determination of $a_4$ and $a_6$ include variations due to the uncertainty in the $C_6$ coefficient, and numerical uncertainty.

We can now compare our new determination of $a_6$ to a value strongly constrained by the position of a Feshbach resonance in d-wave collisions measured in  \cite{d-wave-Fesh}. The node position of the $S_t=6$ wavefunction is adjusted to reproduce the binding energy of the first bound level of the ground molecular potential of Cr$_2$, 22.85~MHz, which is deduced (with an error bar of 0.04~MHz) from the precise measurement of the Feshbach resonance in d-wave at 8.155~Gauss (\cite{d-wave-Fesh}). This measurement determines the scattering length of the ground molecular potential $S_t=6$, $a_6$, with a very good precision, limited by the precision in the knowledge of the $C_6$ coefficient. Varying $C_6$ on the whole confidence interval given in ref \cite{werner}, 663-803~atomic units, makes $a_6$ to vary only between 102.2 to 103.9~a$_0$. This is to our knowledge by far the most precise determination of the $S=6$ scattering length of Cr, and it is in very good agreement with the value deduced from the present analysis of dipolar relaxation as a function of magnetic field.

\section{The influence of incoming higher partial waves in dipolar relaxation}

In a gaussian spherical BEC, the relative wavefunction of any pair of particles is also gaussian and spherical, which results in purely s-wave scattering. Hence, despite the fact that dipolar interactions are long range, dipolar relaxation in a BEC can only be due to an incoming partial wave $l=0$. For cold thermal gases, collisions for short-range interactions are also only in $s-$wave, as, according to the Wigner-threshold law, the collision cross-section in partial wave $l$ varies as $\epsilon^{2l}$, with the collision energy $\epsilon$: a pair of particles colliding in a partial wave $l>0$ cannot approach each other due to the centrifugal barrier. The issue is different in the case of long-range interactions such as dipole-dipole interactions, for which atoms do not need to come close to each other in order to interact. For example non-zero partial waves contribute to  elastic dipole-dipole scattering even at low temperatures. In this part of the paper, we discuss the case of dipolar relaxation, for which we will see that, at low temperature and not too low magnetic fields, contributions to dipolar relaxation due to collisions in non-zero partial waves can be neglected. We will discuss implications of this feature for the case of ultra-cold fermionic dipolar particles.

To investigate the role of higher partial waves in dipolar relaxation, we have repeated the measurements of the dipolar relaxation rate parameter for thermal gases, by operating at different values of the optical trap depth at the end of the evaporation ramp, corresponding to different temperatures of the gas above the critical temperature. For each value of the trap depth, oscillation frequencies in the trap were measured by parametric excitation experiments. After the sample has been cooled down to its final temperature T, we set the magnetic field to a given value $B$, and send an rf sweep pulse to the atoms to flip the spin of all atoms to $m_S=3$ (similar to what was done with BECs). We then record the number of atoms and their temperature as a function of the time delay after this rf sweep. To relate eq \ref{decayn} to the rate equation for the total number of atoms, we now assume Boltzmann thermal equilibrium in a non-interacting gas. Therefore:
\begin{equation} \label{decayNth}
\frac{dN}{dt}=-\beta \frac{n_0}{2^{3/2}} N -\Gamma_1 N
\end{equation}
with $n_0$ the peak density of the cloud. Eq. (\ref{decayNth}) has an analytical solution which we use to fit to the experimental data, to deduce $\beta$.

\begin{figure}
\centering
\includegraphics[width= 2.8 in]{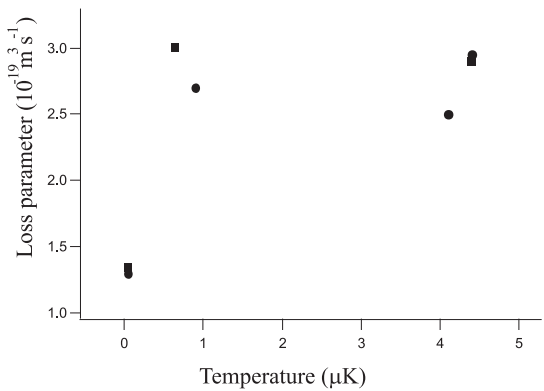}	
\caption{\setlength{\baselineskip}{6pt} {\protect\scriptsize
Filled squares and circles: loss parameter as a function of temperature for two slightly different magnetic fields close to the minimum of dipolar relaxation (filled circles: 3.8 G; filled squares: 3.9 G). The lowest temperature corresponds to a pure BEC, whereas for other data points the gas is fully thermal. The figure shows the factor of 2 reduction of dipolar relaxation when BEC is reached as well as the fact that dipolar relaxation does not depend on temperature for thermal gases.}} \label{temperature}
\end{figure}

We have repeated this procedure for different values of $B$, and the measured thermal loss rate parameter is consistently a factor of roughly 2 higher than the loss rate parameter that we measure for similar magnetic fields for a pure BEC. This factor of 2 corresponds to the factor of two difference between the second order correlation function of a thermal gas and a BEC at short range. This was already calculated in the case of dipolar relaxation of Hydrogen atoms in presence of a BEC in \cite{verhaarstoof}, and we present our own calculation of this phenomenon in the first annexe of this paper.

The results of the measurement of $\beta$ as a function of temperature appear in Fig. \ref{temperature}. At the lowest temperature in this figure the cloud is entirely a BEC, whereas it is completely thermal for the other data points.  Above the critical temperature for BEC, we find the loss rate to be relatively insensitive to temperature, and about factor of two larger than in the BEC. Outside of the dip in dipolar relaxation, and at sufficiently large magnetic fields, we do expect dipolar relaxation to be insensitive to temperature, as shown by eq. (\ref{sigma1born}) and (\ref{sigma2born}). We also show in the first theoretical annexe of this paper that this property remains valid in the dip, as we experimentally demonstrate here.

\begin{figure}
\centering
\includegraphics[width= 2.8 in]{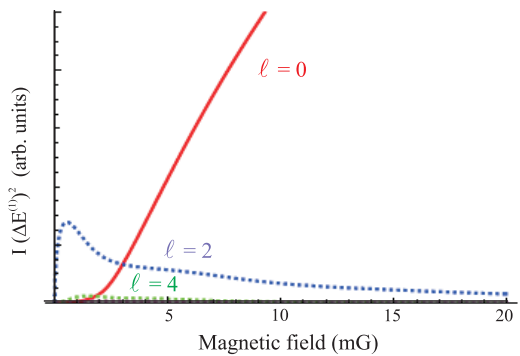}	
\caption{\setlength{\baselineskip}{6pt} {\protect\scriptsize
Calculation of dipolar relaxation to state $|out^{(1)}>$ starting from either $l=0$ (red solid line) or from higher partial waves (blue dotted line: from $l=2$; green dotted line: from $l=4$. Dipolar relaxation through even larger partial waves lead to even smaller contributions)}} \label{partialwaves}
\end{figure}

The data shown in Fig. \ref{temperature} are taken at a magnetic field value corresponding to the minimum in dipolar relaxation shown in Fig. \ref{antiresonance} (3.8G). At this value, the measured value of the loss rate parameter for the thermal gas is a factor of 20 smaller than the loss rate parameter deduced from eq. (\ref{sigma1born}) and (\ref{sigma2born}). As explained above in the case of the BEC, both the position and the magnitude of such dip are fully explained by a theory which only includes $s-wave$ scattering. The similar strong reduction of dipolar relaxation for thermal gases at 3.8 G therefore suggests that other incoming partial waves do not play any role in dipolar relaxation for these cold thermal gases. We have verified by numerical calculations (shown in Fig \ref{partialwaves}) that non zero partial waves have a negligible contribution to dipolar relaxation for magnetic fields larger than 30 mG. The calculation is described in Annexe I.

One may wonder why higher partial waves do not contribute to dipolar relaxation, whereas they do contribute to elastic dipole-dipole interactions.  For dipolar relaxation at relatively high fields, the energy in the output channel is much higher than the initial energy. As a consequence, $\Psi_{out}(r)$ oscillates much faster than  $\Psi_{in}(r)$ at long distances. Most of the contribution of the integral of  $\Psi_{in}(r) \Psi_{out}(r) 1/r^3$ therefore comes from the region where $\Psi_{out}(r)$ does not oscillate rapidly, $i.e.$ from the region of the classical turning point of the corresponding output potential at the output energy, beyond the rotationnal barrier. For large fields, this point is at short distances. If the input channel is $l=0$, then the input wavefunction is constant at short distances, and non-zero. The integral is then also non-zero. If the input channel is a partial wave $l>0$, $\Psi_{in}(r) \propto r^l$ is very small at short interatomic distances, and the integral is consequently very small. As a conclusion, only the $l=0$ partial wave contributes to dipolar relaxation at low collision energy and high magnetic fields.

On the other hand, when the magnetic field is very small, $\Psi_{out}(r)$ and  $\Psi_{in}(r)$ oscillate with similar spatial frequency, and the  integral of $\Psi_{in}(r) \Psi_{out}(r) 1/r^3$ is not negligible, even if the initial partial wave is $l>0$. Non-zero partial waves therefore contribute to either dipolar relaxation at low collision energy and small magnetic fields, or to elastic scattering due to dipole-dipole interactions.

We have thus shown that although dipole-dipole interactions are long-range, dipolar relaxation is a purely $s-$wave phenomenon at large magnetic fields. Dipolar relaxation will therefore be suppressed for polarized fermions at low temperature and large magnetic fields, since two identical fermions collide in odd partial waves. Indeed, the cross section for dipolar relaxation given by eq. (\ref{sigma1born}), (\ref{sigma2born}) for fermions at low temperature and high fields is proportional to $\frac{k_i}{k_f}$. This means that at low temperature, the inelastic loss parameter is  proportional to $T$, and also that operating at high fields is favorable to reduce dipolar relaxation. This feature is much encouraging for two applications: first, it should be possible to produce stable polarized Fermi degenerate gases with strong dipole-dipole interactions, even in a magnetic trap; second, mixtures of fermions in the two lowest spin states should also be stable with respect to dipolar relaxation: dipolar relaxation starting from a pair of atoms in the two lowest spin states colliding in $s-$wave necessarily results in a pair of indistinguishable fermions in an even partial wave, which is excluded by the Pauli principle.

A second important consequence of the fact that dipolar relaxation is dominated by $s-$wave collisions at low temperature is that one can use the measurement of dipolar relaxation versus magnetic field $B$ as a probe for measuring the number of atomic pairs in the gas at a typical distance $R_{DR}(B)$. In practice, one should use numerical simulations to take into account the inner molecular potentials. Measuring the dipolar relaxation loss parameter as a function of magnetic fields may then be a good way to measure the second order correlation function in thermal gases as a function of interparticle distance, in analogy with the Hanbury-Brown and Twiss effect \cite{hbt}. 
In this way, dipolar relaxation could be a probe for correlations up to distances corresponding to magnetic fields at which higher partial waves significantly contribute $i.e$ up to about 100 nm.

\section{Reduction of dipolar relaxation by confinement}

It is often stated that collisions in traps are barely different from collisions for homogeneous systems, and this is most of the times correct. However, scattering may be modified in certain situations of very tight confinement. Modifications of elastic scattering by confinement is typically obtained when the vibrational frequency of the trap is on the same order of magnitude or larger than the binding energy of the last molecular bound state in the collisional channel. For $s-$wave collisions in a harmonic trap, this happens when the harmonic oscillator size of the trap is smaller than the scattering length of the atoms. How confinement modifies elastic scattering has been theoretically investigated in the case of tight isotropic 3D confinement \cite{julienneconfinement}, 1D confinement resulting in quasi 2D gases \cite{petrovprl,petrovpra}, and 2D confinement resulting in quasi-1D gases \cite{confinement}.

To date, the traps with the strongest confinement are created by means of optical lattices. The typical size of the vibrational ground state in the lattice sites is on the order of 50 nm, much larger than typical scattering lengths. A Feshbach resonance is therefore needed to enter the regime where scattering is modified by confinement. Confinement-induced molecules were for example observed for a 1D Fermi gas near a Feshbach resonances \cite{moritz}.

Inelastic scattering may also be modified in situations of strong confinement \cite{petrovprl, petrovpra, LiKrems}. In the limit of large energy release during the inelastic process (corresponding to large magnetic fields in the case of dipolar relaxation), the output channel wavefunction is not modified by confinement. Inelastic losses are then modified through the modification of the incoming wavefunction due to confinement, and a confinement-induced reduction of inelastic losses occurs when the scattering length is comparable to the harmonic oscillator size \cite{petrovpra,chuloss}.

Here, we show that inelastic scattering is also modified by confinement when the kinetic energy in the output channel is smaller than the vibrational frequency of the trap. In the case of dipolar relaxation, as the gain of kinetic energy $\Delta E$ is directly related to the local magnetic field, it is interesting to work in strongly confining geometries and in weak magnetic fields. We then observe a strong reduction of the dipolar relaxation rate parameter, which arises both because strong confinement reduces the density of states at the final energy, and because the inelastic transitions occur at internuclear distances $R^*$ comparable to the harmonic oscillator size of the trap. We emphasize that this reduction is obtained in a regime in which the scattering length of the atoms is much smaller than the harmonic oscillator size, and that it therefore does not rely on the proximity of a Feshbach resonance.

To create very confining traps in one dimension, we focus a retro-reflected single frequency laser at 532 nm (1.5 W from a Verdi laser), with a waist of 40 $\mu$m, on the Cr-BEC. As the standing wave is turned on sufficiently slowly (we use a 30 ms linear voltage ramp controlling an acousto-optical modulator), the BEC is loaded adiabatically in the first band of a 1D optical lattice, whose depth is between 30 and 35 $E_R$, where $E_R = \frac{h^2}{2 m \lambda^2}$ is the recoil energy, with $\lambda=532$ nm. The lattice beams propagate along the vertical axis (Oz). The vibrational spacing $\omega_L / 2 \pi$ in the lattice is on the order of 130 kHz, much larger than any other energy scale in the system (temperature, chemical potential); consequently,  the motion is frozen in one dimension. The size of the harmonic oscillator ground state wave function in the lattice $a_L=\sqrt{\hbar/m \omega_L}$ is about 35 nm. At this lattice depth, the tunneling time from one lattice site to the next is  long (400 ms) compared to the timescale of the experiment (up to 100 ms), so that we consider that the BEC is split into an array of independent 2D BECs. Indeed, time-of-flight images, obtained by letting the atoms expand after a sudden release of the optical lattice, show a gaussian profile along the axis of the lattice, and a Thomas-Fermi profile perpendicular to the lattice axis.

After the BEC is loaded in the optical lattice, we reduce the magnetic field, and set it to a low value, so that the Larmor frequency $\omega_0=g_S \mu_B B$ is chosen from below the vibrational spacing (130 kHz) to slightly below the lattice depth (400 kHz). For all experiments, the magnetic field vector is in the plane of the 2D gases, $i.e.$ perpendicular to the optical lattice beams.  We then flip the spin of the atoms with a radio-frequency sweep as described earlier in this paper. This produces an array of 2D BECs in the $m_S=3$ state. Due to the optical lattice, the peak density of the BEC is raised up to 5 10$^{20}$ m$^{-3}$, noticeably higher than in the 3D BEC.

We then observe a very rapid heating of the cloud, after which we loose the BEC: a small fraction of the atoms undergoes dipolar relaxation; these hot  atoms (their energy is set by $\Delta E$) remain trapped in the optical lattice, and thermalize with the other atoms, so that the temperature of the cloud rapidly exceeds the degeneracy temperature. This process happens within a timescale so short that we have not been able to experimentally analyze its dynamics.

\begin{figure}
\centering
\includegraphics[width= 2.8 in]{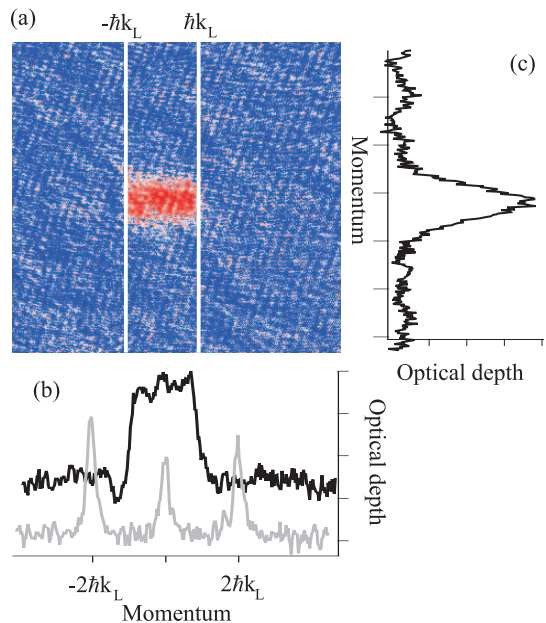}	
\caption{\setlength{\baselineskip}{6pt} {\protect\scriptsize
(a) Absorption picture taken 5 ms after the optical lattice has been released adiabatically (band-mapping procedure). (b) (black) Cut of picture (a) along the axis of the lattice, as compared to (grey) a cross section of Bose-condensed atoms 5 ms after having been submitted to a brief pulse of the optical lattice, thus producing diffraction peaks at $\pm 2 \hbar k_L$: the band-mapping procedure shows that all atoms lie inside the first Brillouin zone, showing no population in higher lattice bands.  (c) Cut of picture (a) perpendicular to the optical lattice axis, from which we deduce the average temperature in the lattice.}} \label{bandmapping}
\end{figure}

Before turning to the measurement of the dipolar relaxation rate parameter for thermal gases in the lattice, we must first check that all particles remain in the lowest band of the lattice. In fact, for most of our experimental situations (when $g_S \mu_B B > \hbar \omega_L /2$, see below), the energy gained by one atom in a dipolar relaxation event is larger than the vibrational spacing in the lattice. Some of the atoms can therefore be excited to higher vibrationnal states in the lattice by dipolar relaxation. We thus operate in a regime where the time $t$ spent in the $m_S=3$ state is small enough that most atoms do not undergo dipolar relaxation. We check that indeed at any given time the vast majority of the atoms are in the lowest energy band, by reducing the lattice depth in 100 $\mu$s, which is slow compared to the timescale for band excitation in the lattice ($10 \mu$s), but rapid compared to the timescale for thermalization (on the order of 1 ms): this band mapping procedure \cite{bandmapping} adiabatically transfers the quasi-momentum distribution in the lattice into the real momentum distribution. The population in the n$^{th}$ Brillouin zone corresponds to the population in the n$^{th}$ band. Typical results are represented fig \ref{bandmapping}. We do observe that (within the signal to noise ratio) almost all atoms are trapped in the lowest band of the lattice. 

One ms after the rf sweep, the system therefore consists of an array of about forty 2D thermal clouds. The typical temperature is 300 nK, and the typical number of atoms in the central site of the lattice is 250. The 3D peak density is $n_0^{3D} =$ 1.5 $10^{20}$ m$^{-3}$, and the 2D peak density is $n_0^{2D} =  10^{13}$ m$^{-2}$, corresponding, in the central site, to a degeneracy parameter  $n_0^{2D} \Lambda^2 \approx 1$, where $\Lambda=\frac{h}{\sqrt{2 \pi m k_B T}}$ is the thermal de Broglie wavelength. The degeneracy parameter is therefore below the critical phase-space density value for the Berezinsky-Kosterlitz-Thouless transition \cite{bkt}, and the 2D gas is thermal. Indeed, when the lattice is switched off abruptly and the atoms let free to expand for a few ms, the profile of the cloud reveals a gaussian velocity distribution perpendicular to the lattice.

\begin{figure}
\centering
\includegraphics[width= 2.8 in]{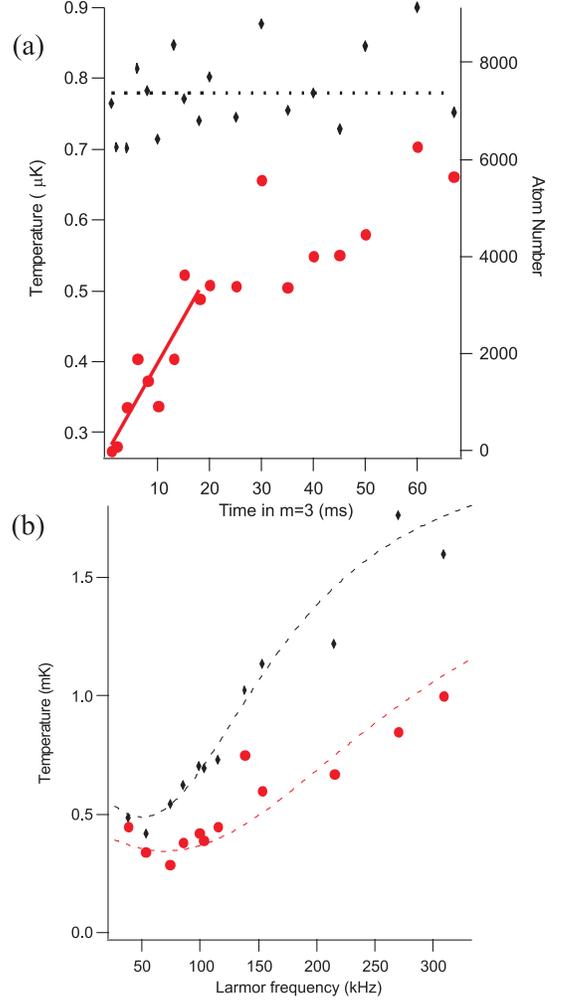}	
\caption{\setlength{\baselineskip}{6pt} {\protect\scriptsize
Heating of the atoms in the lattice due to dipolar relaxation, without losses. (a) Typical data for the number of atoms (black diamonds) and temperature (red bullets) of the cloud as a function of the time spent in $m_S=3$ in the lattice; the lattice depth is 35 $E_R$, the Larmor frequency is $\omega_0=$100 kHz. The straight line is a linear fit to the data used to infer the heating rate. The dotted line is a straight line to guide the eye (b) Temperature of the cloud for $t=0.5$ ms (red bullets) and $t=20$ ms (black diamonds), as a function of $\omega_0$. Dashed lines are guides for the eye.}} \label{heating2D}
\end{figure}

To estimate the dipolar relaxation rate coefficient, we first measure the heating rate in the array of 2D thermal clouds of $m_S=3$ Cr atoms. The temperature of the 2D thermal gases is deduced by using the band mapping procedure explained above, and fitting the momentum distribution perpendicular to the lattice to extract an effective 2D temperature. This temperature is an average over the temperatures in each 2D cloud, weighted by the number of atoms in each 2D cloud. As shown in fig \ref{heating2D} (a), the temperature of the cloud increases linearly at short times, whereas the number of atoms is constant. To deduce the heating rate, we can therefore measure the temperature of the cloud for two different (but short enough) durations spent in $m_S=3$, as shown in fig. \ref{heating2D} (b). This yields a fast measure of the heating rate in the lattice as a function of the magnetic field.

To relate the measured heating rate to a rate parameter for dipolar relaxation, we also need to estimate the density profile of the cloud. We cannot individually address the different lattice sites to measure their number of atoms. The total number of atoms remains unchanged (see fig \ref{heating2D} (a)), and the tunneling time (400 ms) is much longer than the timescale of the experiment (20 ms). We therefore assume that the number of atoms in each 2D thermal gas is identical to the number of atoms in this site once the BEC has been adiabatically loaded in the optical lattice, which we calculate, assuming the loading time is long enough that the chemical potential remains homogeneous across the sample. To estimate the density, we also measure the trapping frequencies by parametric excitation of the cloud with and without the lattice. We calibrate the lattice depth by applying a short pulse of the lattice light to the BEC and analysing the growth of the diffraction peaks as a function of time.

Assuming that the density profile of the cloud along the axis of the optical lattice is identical to the one of the initial $m_S=-3$ BEC loaded into the optical lattice, and assuming a thermal profile perpendicular to the lattice, we can deduce the rate equation for the number of atoms in the $m_S=3$ state ($N_3$) by integrating eq. (\ref{decayn})  over the density profile:
\begin{equation} \label{heating}
\frac{dN_3}{dt}= - \beta \frac{3}{7 \sqrt{2}} (n_0^{3D})_3 N_3 -\Gamma_1 N_3
\end{equation}
where $(n_0^{3D})_3$ is the peak density of $m_S=3$ atoms in the central lattice site. This equation also assumes that most atoms are in the $m_S=3$ state.

\begin{figure}
\centering
\includegraphics[width= 2.8 in]{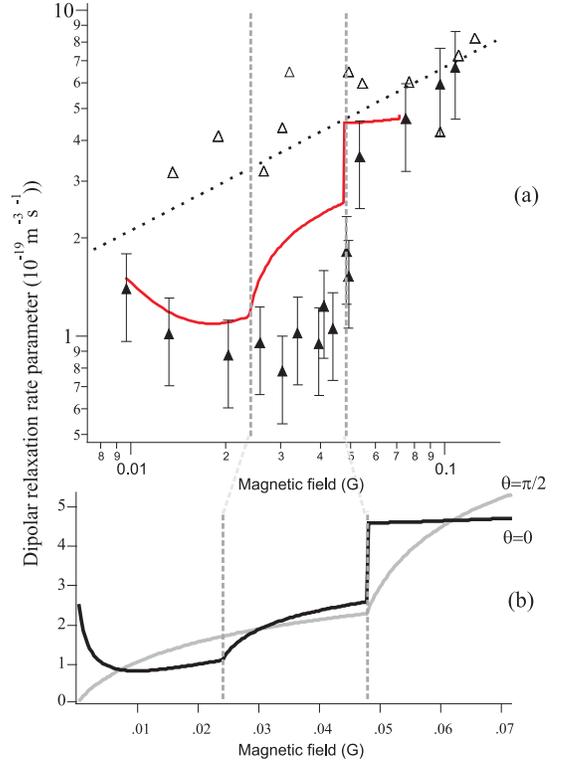}	
\caption{\setlength{\baselineskip}{6pt} {\protect\scriptsize
(a) Dipolar relaxation loss parameter as a function of magnetic field, in the lattice (filled triangles) and without the lattice (open triangles). The dashed vertical lines represent (left) the magnetic field above which excitation by dipolar relaxation of the first vibrational band becomes energetically allowed (first threshold) ; (right) the second threshold. The red solid line is the result of the 2D theory for the specific orientation of the magnetic field in the experiment, and the black dotted line the result of the 3D theory. (b) results of the 2D theory for a magnetic field parallel to the 2D planes (black line) and perpendicular to them (grey line).}} \label{reduceddimension}
\end{figure}

The small decrease of the number of $m_S=3$ atoms does not lead to losses, as $\Delta E^{(j)}$ is smaller than the trap depth. Rather, dipolar relaxation results in heating. The energy gain due to dipolar relaxation is only redistributed in two dimensions, perpendicular to the lattice axis, as shown by figure \ref{bandmapping}. We assume that the trap is harmonic in these two directions, and we use the equipartition theorem to relate the heating rate $dT/dt$ to the rate of increase of energy due to dipolar relaxation:

\begin{eqnarray}
\frac{dE}{dt} = 2 k_B N_3 \frac{dT}{dt} \nonumber \\
= \frac{3}{7 \sqrt{2}} (n_0^{3D})_3 N_3 \left( \beta_1  \Delta E^{(1)} /2 + \beta_2 \Delta E^{(2)} /2 \right)
\label{heatinglatticebeta}
\end{eqnarray}
These equations are valid as long as most atoms remain in the $m_S=3$ state. As $\Delta E ^{(2)} = 2 \Delta E ^{(1)}$, and as both channels may be modified differently by confinement, we define a new dipolar relaxation 'heat' rate parameter $\beta_{heat}^{2D} = \beta_1 /2 + \beta_2$. We then use eq. (\ref{heatinglatticebeta}) to relate the measured heating rate to $\beta_{heat}$. Results of this procedure as a function of the magnetic field are represented in fig. \ref{reduceddimension}.

We also measure for low magnetic fields the dipolar relaxation parameter for a 3D geometry, $i.e.$ for a thermal cloud without the lattice under otherwise similar conditions. For the weakest magnetic fields, dipolar relaxation results in heating of the cloud, which increases evaporation in the trap. We observe a loss rate which we therefore attribute to the sole evaporation process. From the loss rate of atoms, and assuming that the average energy lost per evaporated particle is $(\eta+1) k_B T$ where $\eta$ is the estimated ratio between the trap depth and the temperature \cite{luiten}, and assuming equipartition of the energy in 3D, one can therefore relate the heating rate to $\beta_{heat}^{3D} = \beta_1/2 + \beta_2$  using eq. (\ref{decayNth}):

\begin{equation} \label{heating3D}
3 k_B \frac{dT}{dt}= (g \mu_B B) \frac{1}{2 \sqrt{2}} n_0 (\beta_1/2+ \beta_2) - k_B T \frac{dN}{Ndt}(2-\eta)
\end{equation}.

This formula is valid as long as $\Delta E ^{(j=(1,2))}$ is smaller than the trap depth. For larger trap depths, we simply use eq.  (\ref{decayNth}) to determine $\beta$ from the loss rate of particles. In 3D, we can directly relate the total loss rate parameter $\beta$ to $\beta_{heat}^{3D}$, as in the limit where $k_f >> k_i$, $\beta_2/\beta_1= \sqrt{2}/3$: $\beta_{heat}^{3D} = \frac{1}{14} \left(5+3 \sqrt{2}\right)\beta$.

We present in figure \ref{reduceddimension} (a) the experimental results in 3D and in 2D, $\beta_{heat}^{3D}$ and $\beta_{heat}^{2D}$, along with the result of the calculation for the rate parameter in 3D, corresponding to eq. (\ref{sigma1born}) and (\ref{sigma2born}), and in 2D (see below). In this range of magnetic field strengths, our experimental results in 3D are in good agreement with the theory of \cite{hensler}. In addition, we observe that when the magnetic field is high enough, the measured rate parameter in the lattice is not distinguishable from the 3D case. This is in agreement with predictions of \cite{petrovpra}: when the scattering length is small compared to the size of the harmonic oscillator in the lattice and when the energy of the output channel is large compared to the vibrational energy in the trap, the quasi-2D loss rate parameter is not modified by confinement.

When the magnetic field is lowered, the value of the experimental 2D rate coefficient is gradually reduced compared to the 3D case. We also see in Fig. \ref{reduceddimension} that the functional form of the rate coefficient as a function of magnetic field is different from the 3D model. We therefore identify a clear and strong effect of confinement on the physics of dipolar relaxation in 2D.

To interpret our experimental results, we cannot use the theory previously developed in this paper. The main reason is that, as the magnetic field is reduced, dipolar relaxation takes place at longer and longer interatomic distances; for a magnetic fields $B \approx \hbar \omega_L / g_S \mu_B $, the distance at which dipolar relaxation occurs $R_{DR}$ (see eq. \ref{distancerelaxation}) is of the order of the harmonic oscillator size in the lattice $a_{L}=\sqrt{\frac{\hbar}{m \omega_L}}$. As a consequence, the spherical symmetry of the problem is broken, and the partial wave basis which we have used in 3D is not a good basis anymore: the remaining good quantum number is that associated with the projection of the angular momentum along the axis of the 1D lattice, $m_z$. $R_{DR} \approx a_{L}$ also corresponds to a situation where the energy released in a dipolar relaxation event, $\Delta E ^{(i)}$,  is of the order of $\hbar \omega_L$. We have therefore developed a 2D scattering theory based on the Born approximation, which includes both channels of dipolar relaxation, and describe explicitly the possible vibrational excitation in the lattice, from the initially populated $v=0$ state to the $v=0,1$ and $2$ states. We have not calculated the excitation to higher bands, because this would correspond to magnetic fields larger than the ones experimentally investigated. In this theory, lattice sites are assumed to be independent, and the potential in a given lattice site is approximated to a parabola. Our theoretical model is described in the second annexe of this paper.

Results of the calculation are shown in figure \ref{reduceddimension} (b). The calculation shows two consecutive thresholds as the magnetic field is raised. Indeed, if the magnetic field is low enough, population of the $v=1$ and $v=2$ states through dipolar relaxation is energetically forbidden. When $g_S \mu_B B > \hbar \omega_L/2$, excitation to $v=1$ becomes energetically allowed through channel 2 of dipolar relaxation: the pair of atoms gains an energy larger than the vibrational excitation gap of the motion in the lattice. The calculation hence shows the opening of different dipolar relaxation channels ($v=0$ to $v=1$, then $v=0$ to $v=2$) as the magnetic field is increased. This illustrates the effect of the reduction of the density of states on dipolar relaxation as confinement is increased.

As also shown in figure \ref{reduceddimension} (b), dipolar relaxation in 2D strongly depends on the magnetic field orientation $\theta$ with respect to the 2D planes of the 1D lattice sites ($\theta=0$ corresponds to $\vec B$ parallel to the planes). This is a direct consequence of the breakdown of spherical symmetry, and the result of an interplay between the asymmetry of the trap and the anisotropy of dipolar relaxation.

In our experiment, the applied magnetic field is parallel to the 2D plane of the 1D lattice sites. However, for the lowest magnetic fields, our knowledge upon the orientation of the total magnetic field is reduced, because a stray magnetic field perpendicular to the plane may remain. We therefore plot in figure \ref{reduceddimension} (a), along with the experimental data, calculations corresponding to a magnetic field consisting of a combination of a small perpendicular magnetic field and of a parallel magnetic field (the one we apply), which largely dominates for $B>0.025$ G. The agreement between the experimental data and the theory is then good, and best for a remaining perpendicular magnetic field of $B_{perp}=0.01$ G. There seems to be a small systematic shift between experiment and theory, which may be a sign that we overestimate the density in the lattice.

An interesting feature in figure \ref{reduceddimension} is that as the magnetic field is reduced below $\hbar \omega_L/2$, instead of decreasing as one would naively guess, the dipolar relaxation rate parameter increases again. As shown by our theoretical calculation, this phenomenon depends on the  orientation of the magnetic field relative to the plane defined by $\theta$. Indeed, in the zero magnetic field limit, dipolar relaxation vanishes for $\theta =\pi/2$, whereas it reaches a fixed value for  $\theta =0$. This is because in the first case, an $m_z=0$ collisional wave is only coupled to $m_z=2$, and the overlap between these two waves gets smaller and smaller when the magnetic field is reduced, because the centrifugal barrier in $m_z=2$ strongly reduces the amplitude of the wavefunction at small interparticle distances. On the other hand, when $\theta =0$, $m_z=0$ is coupled to itself by dipolar relaxation, so that the coupling does not vanish when $B$ goes to zero. Moreover, coupling of $m_z=0$ to itself by dipolar relaxation increases as the magnetic field is lowered, because the radial parts of the corresponding wavefunctions get more and more similar as the kinetic energy in the output wave gets closer and closer to the initial kinetic energy. Interestingly, the value of $\theta$ at which the 2D dipolar relaxation rate parameter is maximum at $B=0$ is different from $\theta=0$. The non-zero value of the dipolar relaxation rate parameter as $B$ goes to zero, as well as the discontinuity when the second threshold is reached, are both discussed in Annexe II. 

The fact that dipolar relaxation reaches a finite value at $B=0$ (and $T=0$) is in sharp contrast to the situation in 3D at $T=0$. This difference comes from the fact that even at the lowest temperatures, the lattice confinement creates momenta (along the lattice axis) which contribute to dipolar relaxation. Hence, the order of magnitude of the 2D dipolar relaxation rate parameter at zero magnetic field is of the order of the 3D rate parameter at zero field for an incoming wavevector $k_i = 1/a_L$ (see also Annexe II). This also means that, although dipolar relaxation is in general smaller in 2D than in 3D, the situation is reversed when the magnetic field is exceedingly small (even smaller than what we experimentally investigate here).

We have thus shown theoretically and experimentally a strong modification of dipolar relaxation due to confinement at low magnetic fields. We observe an increase of dipolar relaxation when $B$ tends to zero below the excitation gap from $v=0$ to $v=1$, in agreement with our theoretical model. In the limit where $B$ tends to $0$, dipolar relaxation in 2D should therefore be more efficient than in 3D for certain orientations $\theta$ of the magnetic field. However, in general, dipolar relaxation is reduced in 2D compared to the 3D case. We observe the opening of new dipolar relaxation channels when the magnetic field is raised, corresponding to the threshold for excitation to $v=1$ and $v=2$. For the largest magnetic fields, the 2D dipolar relaxation rate parameter gets closer to the 3D one. We observe a crossover between a situation ($g \mu_B B < \hbar \omega_L$) for which population of the higher bands is completely excluded based on conservation energy arguments (dipolar relaxation happens in reduced dimensions) to a situation in which the excitation of motion along the third direction of space becomes possible. Then dipolar relaxation happens in 3D, and inelastic scattering is not modified since the scattering length is small compared to the harmonic oscillator size of the trap \cite{petrovpra}.

The confinement-induced modification of dipolar relaxation arises from the combined modification of the input and the output wave functions and is significant when the magnetic field is small enough. In contrast to all other predicted \cite{julienneconfinement,petrovprl,petrovpra,confinement,LiKrems} and observed \cite{chuloss} confinement-induced modifications of elastic or inelastic scattering the modification of inelastic scattering by confinement at low field does not rely on a scattering length comparable to the typical trap size, and can be observed without the use of a Feshbach resonance. This provides a general route to reduce dipolar losses and increase the stability of dipolar (spinor) gases. Our work on dipolar relaxation in optical lattice will soon continue by the study of dipolar relaxation in 2D optical lattices, where dipolar relaxation should completely vanish when the magnetic field is smaller than the excitation energy gap to the first band in the 2D lattice, and parallel to the 1D gases produced by the 2D lattice. Furthermore, in a cubic 3D optical lattice, dipolar relaxation should occur only when the Larmor frequency exactly matches the excitation gap in the lattice. Working in a lattice is then a good way to observe the Einstein-de-Haas effect with cold atoms, as the excitation gap in the lattice reduces the experimental complications linked to controlling the magnetic field to a value very close to zero.  Finally, let us mention the possibility, not included in our theoretical model, of inter-site dipolar relaxation at extremely low magnetic fields (i.e. when $R_{DR} \approx \lambda$).

\section{Control of dipolar relaxation by rf fields}

We have shown above that dipolar relaxation can be controlled by means of a static magnetic field, and by means of confinement. We now turn to experiments where we show that we can also control the strength of the matrix element for dipolar relaxation, as well as the energy of the output channel, by using rf oscillating  magnetic fields. Controlling the energy of the output channel is particularly appealing, as it allows for the study of the modification of dipolar relaxation as the process is changed from exoenergetic and dissipative to coherent, with no release of kinetic energy, without the need to control accurately a magnetic field intensity close to zero. As this resonant regime is reached, description in terms of a Fermi golden rule is expected to break down, and in the case of a BEC, a meanfield picture suggests the coherent production of structures similar to vortices, an effect which bears similarities with the Einstein-de-Haas effect \cite{edh}.

\begin{figure}
\centering
\includegraphics[width= 2.8 in]{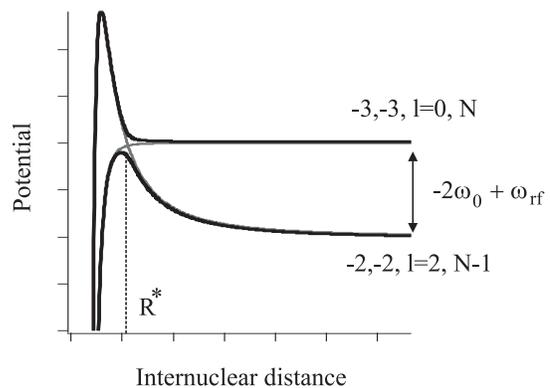}
\caption{\setlength{\baselineskip}{6pt} {\protect\scriptsize
Dipolar relaxation in the presence of rf from the molecular physics point of view, illustrated for channel 2 (sketch); the difference of energy between initial and final channels is controlled by a combination of the Larmor frequency and of the rf frequency and $R^*$ is modified accordingly.}}
\label{principerf}
\end{figure}

Our idea is to use radio-frequency (rf) magnetic fields to dress atoms in the $m_S=-3$ state. Then, without rf, dipolar relaxation is energetically forbidden at low temperature. With rf, we show that dipolar relaxation can occur between states of different manifolds (a process shown in Figure \ref{principerf}). The amplitude of this process is proportional to the dipole-dipole interaction matrix element, and depends on both the frequency and the Rabi frequency of the rf field. In addition, the energy of the output channel depends on the frequency of the rf field. Dipolar relaxation between rf manifolds was to our knowledge first discussed theoretically in \cite{verhaar}, already observed in our experiment as briefly discussed in \cite{beaufilslande}, in a situation where the rf field was perpendicular to the static magnetic field. Here we study the configuration where the rf field is parallel to the static field and we provide analytical formula for rf-assisted dipolar relaxation. The present work also has similarities with previous results in our group describing rf-association in the vicinity of a Feshbach resonance as an assisted Feshbach resonance \cite{beaufilsrf}, or with earlier work with Rydberg atoms \cite{pilletrf}.  

There is an analytical expression for the internal eigenstates of atoms dressed by rf, when the rf field is parallel to the static field, for an arbitrarily large Rabi frequency. As the rf field is parallel to the static field, there is no precession, and the dressed eigenstates are products of the internal Zeeman substate $\left|m_S\right\rangle$, and of a translation operator acting on the rf field. Dressed eigenstates group in a series of manifolds, whose spin structure is identical to that of undressed atoms. Inside a given manifold, the energy of the dressed eigenstate are simply $m_S \hbar \omega_0$ (where $\omega_0$ is the Larmor frequency), and consecutive manifolds differ by the energy of an rf photon $\hbar \omega_{rf}$. The key point is the following: starting from the lowest state of energy of a given manifold (corresponding in the case of chromium to $m_S=-3$), it is now energetically possible to decay into a state of even lower energy, in a lower manifold. In fact, as the spin nature of the dressed eigenstates is not modified by rf, the matrix element between one magnetic state in a manifold and another magnetic state in another manifold is proportional to the matrix element between these two magnetic states without rf. The coefficient of proportionality is given by the overlap between the rf field components of the dressed states corresponding to the input and output channels.

We consider two atoms in the $\left|m_S=-3\right\rangle$ state. The dipolar relaxation reaction 
\begin{equation}
\left|-3,-3\right\rangle \longrightarrow \frac{1}{\sqrt{2}} \left(\left|-3,-2\right\rangle+\left|-2,-3\right\rangle\right),
\end{equation}
for which the matrix element is similar to $V_1$ (see eq. (\ref{matrixc1})), is energetically forbidden at low temperature. However, if one considers this process in presence of rf photons, it becomes energetically allowed, provided it procedes together with the absorption of a sufficiently large number of rf photons $\Delta N >\omega_0/\omega_{rf}$. Starting from the $\left|m_S=-3\right\rangle$ state, in presence of rf, the two different dipolar relaxation channels 1 and 2 correspond to two series of channels for dipolar relaxation triggered by the absorption of $N$ rf photons:

The series of channels corresponding to channel 1 is:
\begin{eqnarray}
\left|-3,-3,N\right\rangle \longrightarrow \frac{1}{\sqrt{2}} \left(\left|-3,-2,N'\right\rangle+\left|-2,-3,N'\right\rangle\right) \\
\label{channel1-}
\Delta E^{(1)} = g \mu_B B + (N'-N) \hbar \omega_{rf} \label{DEC1RF} \\ \nonumber
\end{eqnarray}
and for channel 2 :
\begin{eqnarray}
\left|-3,-3,N\right\rangle \longrightarrow \left|-2,-2,N'\right\rangle \\
\label{channel2-}
\Delta E^{(2)} = 2 g \mu_B B  + (N'-N) \hbar \omega_{rf} \label{DEC2RF} \\ \nonumber
\end{eqnarray}

We now turn to the calculation of the dipolar relaxation rate parameter between these manifolds. In absence of dipole-dipole interaction, the hamiltonian describing two particles in presence of rf reads:

\begin{eqnarray}
H_0&=&\frac{\hbar^2}{2m}\Delta+ H_{rf}+g\mu_BB_zS_z\\
\end{eqnarray}
where $H_{rf}= \hbar\omega a^+a+\lambda S_z(a+a^+)$ is the Hamiltonian for the radio frequency coupling with $2 \lambda \sqrt{N} = \hbar \Omega$ and $\Omega$ the Rabi frequency. The key point is that, since the static magnetic field is parallel to the rf field, the diagonalization of $H_0$ is exact for arbitrarily large rf power, and the eigenstates appear as a series of manifolds of dressed states \cite{cohen}:

\begin{equation}
\left|\widetilde{X,N}\right\rangle = T_X^+ \left|X,N\right\rangle
\label{dressedstates}
\end{equation}
$X$ denotes the internal state of the pair of atoms. $T_X= \exp\left(-\frac{M_X \lambda}{\hbar \omega}\left(a-a^+\right)\right)$ is a field translation operator, with $M_X$ the spin projection of state $X$ along the axis $z$. The  eigenenergies of the dressed states $\left|\widetilde{X,N}\right\rangle$ are $W_X=E_X + M_X \hbar \omega_0 + N \hbar \omega - \frac{M_X^2 \lambda^2}{\hbar \omega}$, where $E_X$ is the eigen-energy of state $X$ without rf or magnetic field. The term $\frac{M_X^2 \lambda^2}{\hbar \omega} = \frac{M_X^2 \Omega}{4 \omega_{rf}} \frac{\Omega}{N}$ is negligible compared to $\omega_{rf}$ since for our typical rf power, $\Omega \approx \omega_{rf}$, and $N>>1$.

In the spirit of the calculation presented in the first part of this paper, without rf, leading to eq. \ref{sigma1born} and \ref{sigma2born}, we first neglect all molecular potentials.  The dressed eigenvectors are the products of three components: a plane wave (for the spatial wavefunction); the internal wave vector of each atom; the (spin dependent) translation operator applied to the rf field state:
\begin{eqnarray}
&\langle \vec r|\vec k,\widetilde{m_1,m_2,N}\rangle= \nonumber \\
&\exp(i\vec k \cdot \vec r)\exp\left[\lambda (m_{1}+m_{2})(a-a^+)\right] |N,m_1,m_2\rangle
\end{eqnarray}
and the eigenvalues are
\begin{eqnarray}
&E_{\vec k, m_1, m_2, N}= \nonumber \\
&\frac{\hbar^2k^2}{2m}+N\hbar\omega -\frac{\lambda^2(m_1+m_2)^2}{\hbar\omega}+g\mu_BB_z(m_1+m_2)
\end{eqnarray}

Using first order perturbation theory, we now calculate the coupling matrix element between two different dressed states by dipole-dipole interactions:
\begin{eqnarray}
A &=& \langle \vec k',\widetilde{m_1',m_2',N'}|V_{\rm dd} |\vec k,\widetilde{m_1,m_2,N}\rangle \nonumber \\
 &=& \langle \vec k',m_1',m_2',N'|T V_{\rm dd} T^+|\vec k,m_1,m_2,N\rangle \nonumber \\
 &=& \langle \vec k',m_1',m_2'|V_{\rm dd} |\vec k,m_1,m_2\rangle  \nonumber \\
&\times& \langle N'|\exp\left(\frac{\lambda \Delta m}{\hbar\omega}(a^+-a) \right)|N\rangle
\end{eqnarray}
where $\Delta m=m'_1+m'_2-m_1-m_2$. The first part of the latter equation is identical to the one describing dipolar relaxation without rf: rf cannot induce coupling between internal states that are not otherwise coupled by dipole-dipole interactions. The idea of the calculation for the second part of this equation is to expand the exponential in series, then recognize the series expansion of Bessel functions when the photon number is very large. We thus obtain:

\begin{eqnarray}
\langle N'|\exp\left(\frac{\lambda \Delta m}{\hbar\omega}(a^+-a) \right)|N\rangle \\
= J_{N-N'}\left(\frac{\Omega_1\Delta m}{\omega}\right)
\end{eqnarray}
which generalizes earlier results which can be found in  \cite{cohen}. The final result of our calculation is therefore simply related to that given in \cite{hensler}:
\begin{eqnarray}
\sigma_1&=&\frac{8\pi}{15}S^3\alpha^2f(k_f^{(1)}/k_i)k_f^{(1)}/k_i|J_{N-N'}(\Omega_1/\omega)|^2 \label{sigma1rf} \\
\sigma_2&=&\frac{8\pi}{15}S^2\alpha^2f(k_f^{(2)}/k_i)k_f^{(2)}/k_i|J_{N-N'}(2\Omega_1/\omega)|^2 \label{sigma2rf} \\ \nonumber
\end{eqnarray}
with $\alpha = \left(\frac{d^2 m}{\hbar^2}\right)$. Like in the situation with no rf, $k_i$ and $k_f$ are related through the conservation of energy equation:
\begin{equation}
\frac{\hbar^2 (k_f^{(j)})^2}{m}=\frac{\hbar^2 k_i^2}{m}+\hbar\omega(N-N')-\mu_B B \Delta m
\end{equation}

\begin{figure}
\centering
\includegraphics[width= 2.8 in]{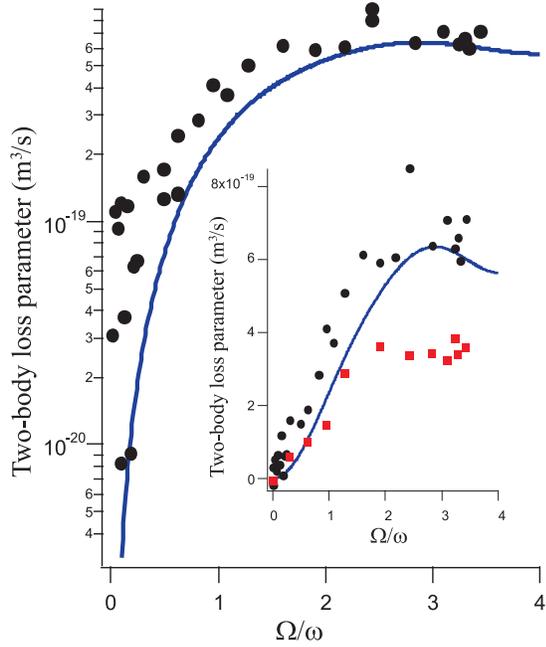}	
\caption{\setlength{\baselineskip}{6pt} {\protect\scriptsize
rf-assisted dipolar relaxation loss parameter as a function of the Rabi frequency, for a thermal gas, in Log scale. Inset (in linear scale): rf-assisted dipolar relaxation loss parameter as a function of the Rabi frequency, for a thermal gas (black filled circles) and for a BEC (red squares). The measured loss rate parameter is roughly a factor of two smaller for a BEC than for a thermal gas. Solid line: theoretical results for a thermal gas.}} \label{rflosses}
\end{figure}

In order to obtain the final result one has to sum over all possible channels, $i.e.$ over all final photon number states and final internal states. The result of this summation is given, with no free parameter, in fig. \ref{rflosses}, along with experimental data. Because we measure loss parameters, and not directly cross-sections, we do not directly plot cross section in figure \ref{rflosses}, but the corresponding loss rate parameter $\beta = 2 \sigma k_i \frac{\hbar}{m/2}$. As the rf frequency is much larger than the initial kinetic energy of the atoms, we  also set $f(k_f/k_i)=2$.

To measure the rf-assisted dipolar relaxation rate coefficient, we first produce either a $m_S=-3$ BEC with the same experimental procedure as above, or a thermal gas just above the critical temperature for BEC (by interrupting the evaporation ramp just before BEC). We then recompress the optical trap (as above, to compensate for gravity without applying any vertical magnetic field gradient), and apply a weak magnetic field, either vertical or horizontal. We then apply a vertically polarized rf pulse, which is short compared to the lifetime of the BEC (a few seconds), but long enough to induce losses. Then we deduce the rf-assisted loss parameter by comparing the number of atoms with and without rf, and taking into account the density distribution in the trap, using eq. (\ref{decayN}) for a BEC or eq. (\ref{decayNth}) for a thermal gas.

Typical results as a function of rf power are presented in fig \ref{rflosses}, for a thermal cloud and for a BEC. For the data of this figure, the rf frequency is set to 300 kHz and the rf field is parallel to the static field. As already observed in \cite{beaufilslande}, the loss parameter saturates when $\Omega / \omega_{rf} \geq 1.5$. Experimental results for the thermal gas are compared to the predictions of eq \ref{sigma1rf}, \ref{sigma2rf}, and we find a very good agreement, with no free parameter. (The rf Rabi frequency is independently calibrated by measuring on-resonance Rabi oscillations.) To emphasize the overall good agreement between theory and experiment, over more than two orders of magnitude in the loss parameter, we plot both the experimental data and theoretical results in Log scale for the thermal gas. As the inset shows, the agreement between theory and experiment is however not perfect. A disagreement of about 20 percent is consistent with our systematic uncertainties (combined uncertainties in trapping frequencies and number of atoms). We also observe that the loss parameter in the BEC is systematically a factor of two lower than in the thermal case, as expected (see above).

\begin{figure}
\centering
\includegraphics[width= 2.8 in]{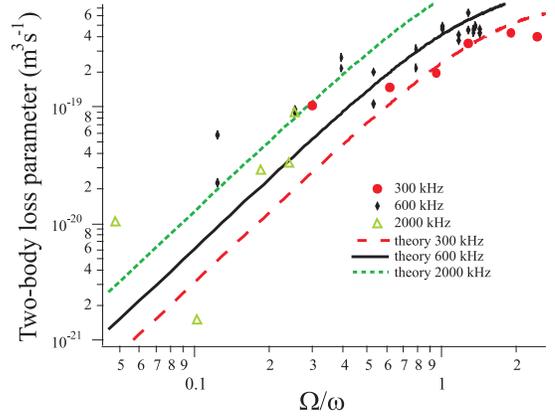}	
\caption{\setlength{\baselineskip}{6pt} {\protect\scriptsize
rf-assisted dipolar relaxation loss parameter for three different rf frequencies (symbols: experimental data; lines: theory). }} \label{rfsyst}
\end{figure}

In our theoretical model, based on the Fermi golden rule, the loss-rate parameter is proportional to the density of states at the output energy, as given by eq. (\ref{DEC1RF}) and (\ref{DEC2RF}). As the output energy is set by the relative value of the Larmor frequency and the rf frequency, this provides a means to experimentally control it. Although we have not directly measured the output energy of the rf-assisted losses, we have performed systematic studies of the rf-assisted dipolar relaxation loss parameter as a function of the rf power for different rf frequencies, all being much larger than the Larmor frequency. As shown in figure \ref{rfsyst}, we find a reasonnably good agreement between theory and experimental data, for all rf frequencies. (In fig \ref{rfsyst}, the theoretical predictions are scaled down by a factor of two to take into account the fact that the data were taken with a BEC.) We hence do verify that losses increase as the rf frequency gets larger, due to the fact that the density of states in the output channel increases with energy.

\begin{figure}
\centering
\includegraphics[width= 2.8 in]{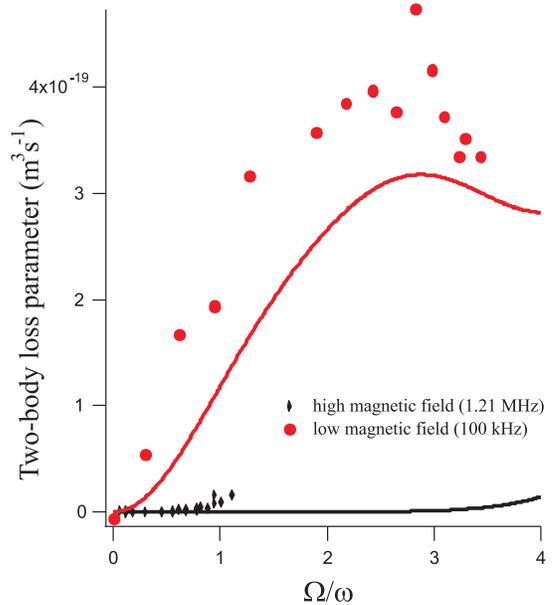}	
\caption{\setlength{\baselineskip}{6pt} {\protect\scriptsize
rf-assisted dipolar relaxation loss parameter as a function of Rabi frequency, for two different values of the bias field (circles: $\omega_0=$100 kHz ; diamonds: $\omega_0=$1.21 MHz). Losses are much suppressed when the rf frequency is much smaller than $\omega_0$, indicating that rf-assisted losses with the absorption of a small ($<5$ in the present case) number of rf photons becomes energetically impossible. Solid lines : corresponding theoretical results}} \label{lowhighfieldrf}
\end{figure}

To further check our ability to control the energy of the output channel by modifying the values of the Larmor frequency and of the rf frequency, we have performed experiments where the rf field is parallel to the static field, and the rf frequency is smaller than the Larmor frequency. Starting with a BEC of $m_S=-3$ atoms, it is then energetically impossible to trigger rf-assisted dipolar relaxation with the absorption of only one rf photon. We performed experiments with an rf frequency of 300 kHz in a static field of .43 G, corresponding to a larmor frequency of $\omega_0=$1.21 MHz; then, at least five rf photons need to be absorbed to provide enough energy for dipolar relaxation. We measured the rf-assisted loss parameter, and our results are represented in Fig \ref{lowhighfieldrf}. In this figure, we compare the experimental measurements for a low ($\omega_0=$105 kHz) and a high ($\omega_0=$1.21 MHz) static field, to the theoretical prediction summing eq. (\ref{sigma1rf}) and (\ref{sigma2rf}) over all energetically allowed rf-assisted dipolar relaxation channels ($i.e.$ $N-N' > \Delta m g_S \mu_B B /\hbar \omega$).

The experiment shows that indeed losses are greatly suppressed in a large static field. This is because dipolar relaxation is energetically forbidden with the absorption of less than 5 rf photons, and because at low rf power, the Bessel functions corresponding to dipolar relaxation with the absorption of 5 or more photons (see eq. (\ref{sigma1rf}) and (\ref{sigma2rf})) are almost equal to zero. We do observe losses when the rf power gets larger $(\Omega/\omega >1$), and they are larger than what our theory predicts. We interpret this disagreement by the following argument: the rf field is not exactly parallel to the static field. Therefore, rf can induce spin precession; the small fraction of the atoms transfered in $m_S=-2$ can undergo dipolar relaxation (within their own manifold), and thus be ejected from the trap.

We have thus shown that dipolar relaxation with the absorption of one photon can be suppressed when the rf photon frequency is smaller than the Larmor frequency, because such dipolar relaxation is then energetically forbidden. This is an indirect signature that the output energy of a dipolar relaxation event is indeed set by the difference of energy between the rf frequency and the Larmor frequency. Future studies will show whether a good control of the output energy, by varying the rf frequency for a fixed bias magnetic field value, is possible or not. This would have the advantage of allowing the analysis of dipolar relaxation in a resonant regime, with the possible observation of the Einstein-de-Haas effect \cite{edh}, easing the requirement of a very good control of the magnetic field at very small values.

\begin{figure}
\centering
\includegraphics[width= 2.8 in]{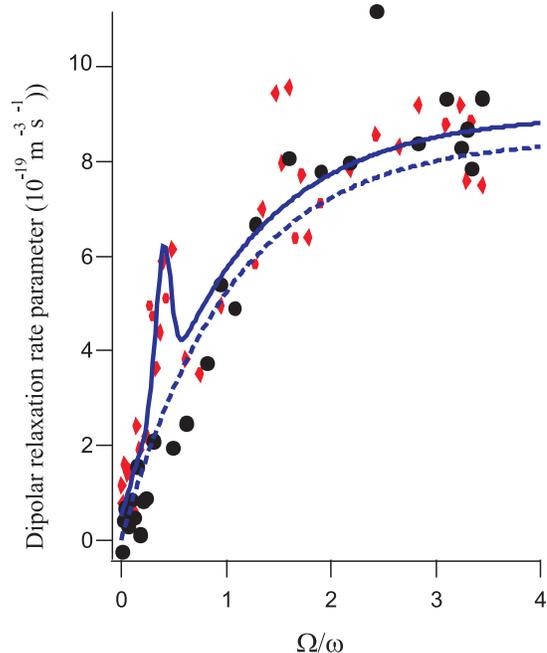}	
\caption{\setlength{\baselineskip}{6pt} {\protect\scriptsize
rf-assisted dipolar relaxation loss parameter for two different orientation of a low static magnetic field (black circles: static field parallel to rf field; red bullets: static field perpendicular to rf field); lines are guides for the eye.}} \label{perp}
\end{figure}

As a complementary information, we compare the measured rf-assisted dipolar relaxation loss parameter when the rf is parallel or orthogonal to a small bias field in figure \ref{perp}. In the case where the static magnetic field is perpendicular to the rf field, we observe an additional resonant loss peak (in the present case for $\Omega/\omega \approx 0.4$). We have attributed this resonance as being due to technical noise in the rf amplifier that we use, which produces heating and losses when a peak of the noise spectrum of the amplifier corresponds to the (Rabi frequency dependent, see \cite{beaufilslande}) Larmor frequency of the atoms.

Overall, apart from the aforementioned loss peak due to technical noise, we see no noticeable difference whether the rf is parallel or perpendicular to the small bias field  (see fig. \ref{perp}). As we operate in a regime where the bias field is much smaller than the rf field, this result is natural: in the low bias field limit, the orientation of the field does not matter. However, we stress that in general the relative orientation of the rf field and the bias field does matter. In general, when the bias field is not very small and the rf field is perpendicular, the loss rate parameter is different from the results of eq. (\ref{sigma1rf}) and (\ref{sigma2rf}), and even the mecanism for dipolar relaxation is modified. In particular, as for example shown in \cite{beaufilslande}, losses correspond to dipolar relaxation between manifolds, but without changing the total magnetic number of the pair of particles. Therefore, the energy of the output channel is in the perpendicular case completely set by the rf frequency.

Hence, we have shown that dipolar relaxation can be engineered by rf fields, and that the rate parameter can be controlled by tuning the Rabi frequency of the field. In the case where the rf field is parallel to the bias field, the energy of the output channel is set by the difference between the rf frequency and the Larmor frequency (for channel 1), and by the difference between the rf frequency and twice the Larmor frequency (for channel 2). As it is technically very difficult to control a magnetic field close to the zero value at the level required to observed the Einstein-de-Haas effect with cold atoms \cite{edh}, we suggest that one could use rf-assisted dipolar relaxation for this goal: indeed, controlling both a static field to about 1 kHz in a given direction and a rf frequency to the same accuracy is a relatively easy experimental routine. The use of low rf power $\Omega/\omega <0.1$ insures that dipolar relaxation with the absorption of only 1 rf photon dominates all other loss mechanisms, as $J_1(\Omega/\omega)$ is then much higher than $J_{l}(\Omega/\omega)$ with $l>1$. A very favorable effect is that the modification of the Larmor frequency due to stray transverse magnetic fields is then reduced by a factor  $J_1(\Omega/\omega)$. We will work towards this goal in future experimental studies.

\section{Conclusion}

To summarize, we have described and experimentally characterized various means to control dipolar relaxation using external fields. Let us emphasize  five major points:

(i) We have discovered a range a magnetic fields in which dipolar relaxation is reduced by up to a factor of 20 compared to previous theoretical predictions \cite{hensler}. 

(ii) We have used this effect for a new accurate determination of the $S=6$ scattering length of Cr atoms. Our result is in very good agreement with another new result presented here, obtained by the analysis of a Feshbach resonance in $d$-wave collisions \cite{d-wave-Fesh}. Both results reduce the uncertainty in $a_6$ from about 15 percent \cite{werner} down to the percent level.

(iii) We have demonstrated that dipolar relaxation is purely $s-$wave at low temperature and high magnetic fields. This is in contrast to elastic dipole-dipole interactions, in which all incoming partial waves contribute.

(iv) We have shown that dipolar relaxation can be modified in 1D optical lattices producing independent 2D gases. The modification by confinement of both the initial collision wavefunction and the final one leads to changes in dipolar relaxation, most significant when the magnetic field is small. Then dipolar relaxation depends on the relative orientation of the 2D gases and the magnetic field, and gaps in the excitation spectrum generally lead to a reduction of dipolar relaxation. A reduction of a factor of up to 7 was observed.

(v) Finally, we have demonstrated rf-assisted dipolar relaxation, which provides a way to control both the dipolar relaxation rate parameter and the energy in the output channel.

Let us finally show how the results presented here allow to suggest possible future experiments.  

(i) It should be possible to use the dip in dipolar relaxation for other atoms (or molecules) with large enough dipolar relaxation rates , for example as a new means to measure scattering lengths, or simply to reduce losses.

(ii) One could use the selective cancelation of the second channel of dipolar relaxation and use dipolar relaxation in the other channel to create a source of pairs of correlated atoms in state $\frac{1}{\sqrt{2}} \left(\left|3,2\right\rangle+\left|2,3\right\rangle\right)$ with orbital momentum $l=2$.

(iii) Since only the $s-$wave collision incoming channel contributes to dipolar relaxation at low temperature, dipolar relaxation should be strongly suppressed for polarized ultra-cold fermions, despite the presence of elastic dipole-dipole interactions: dipolar Fermi gases should be stable in magnetic traps, as well as some trapped mixtures of dipolar fermions.

(iv) As dipolar relaxation takes place at a specific, magnetic-field dependent, interatomic distance, one could use it as a probe of correlation functions in either thermal or quantum degenerate samples.

(v) Strong confinement modifies and generally inhibits dipolar relaxation. Going to 2D or 3D optical lattices, dipolar relaxation should vanish below a certain magnetic field threshold. In particular, by loading several atoms into each site of a 3D optical lattice, one should completely suppress dipolar relaxation, except when the energy released in a dipolar relaxation event exactly matches the energy of one excited band in the lattice. This proposal is promising in the prospect of observing with cold atoms the equivalent of the Einstein-de-Haas effect.

(vi) Rf fields provide a way to control dipolar relaxation, and the output energy. This fact (possibly combined with the control provided by strong confinement) could also be an interesting experimental avenue for the observation of the Einstein-de-Haas effect in a gas of ultra-cold atoms.

Acknowledgements: LPL is Unit\'e Mixte (UMR 7538) of CNRS and of Universit\'e Paris Nord. We acknowledge financial support from Minist\`{e}re de l'Enseignement Sup\'{e}rieur et de la Recherche (CPER), IFRAF (Institut Francilien de Recherche sur les Atomes Froids) and from the Plan-Pluri Formation (PPF) devoted to the manipulation of cold atoms by powerful lasers. P. Pedri and G. Bismut acknowledge financial support by IFRAF.

\section{Annexe I: Calculation of the dipolar relaxation rate:
contribution of initial $\ell$-waves with $\ell \ne 0$ and temperature-dependence }

This theoretical annexe describes dipolar relaxation from the many-body point of view. We use this approach to interpret three experimental results detailed in the paper: the dipolar relaxation rate parameter is twice higher in a thermal gas than in a BEC; dipolar relaxation does not depend on temperature at large magnetic field; at large magnetic fields, only $s-$ wave contributes to dipolar relaxation.

We describe the initial state of the N indistinguishable bosonic atoms ($m_S=3$ Cr atoms) by a fully symmetric state in the occupation number representation:
\begin{equation}
|In>=|n_{\alpha _i}>=\prod_i \frac{\left(a_{\alpha_i}^{\dag}\right)^{n_{\alpha_i}}}{\sqrt{n_{\alpha_i}!}} \left|vac\right\rangle,
\label{nboccup}
\end{equation}
where the $\alpha_i$'s represent the different vibrational states in the trap. $a_{\alpha_i}^{\dag}$ is the particle creation operator relative to the single particle state $\alpha_i$. $\left|vac\right\rangle$ is the vacuum state. We consider two simple situations: a pure BEC, with $n_{\alpha_0}=N$ and all other occupation numbers equal to zero, and a pure thermal cloud in an isotropic harmonic trap of frequency $\omega /(2\pi)$; the occupation numbers then follow the Maxwell-Boltzmann statistics, with
\begin{equation}
n_{\alpha_i}=N\exp(-\beta E_{\alpha_i})/\sum_i \exp(-\beta E_{\alpha_i}),
\label{nboccuptherm}
\end{equation}
where $\beta = 1/k_B T$, $k_B$ being the Boltzmann constant and $T$ the temperature.
In the second-quantization formalism, the dipole-dipole interaction operator reads
\begin{equation}
V=\frac{1}{2} \sum_{\xi, \eta,\lambda, \zeta} a^{\dag}_{\xi} a^{\dag}_{\eta} a_{\lambda} a_{\zeta}
<\xi \eta |V_{dd}| \lambda \zeta >,
\label{sdquant}
\end{equation}
where the sum runs over all atomic trapped and untrapped states and where the matrix element is a non-symmetrized two-atom matrix element.

In the final state of a dipolar relaxation event, two non-zero occupation numbers of the states $\alpha_i$ ($i=n,m$) which are initially populated (trapped states) will decrease by one (or one occupation number will decrease by two), whereas two new (untrapped) states $\gamma_i$ ($i=(p,q)$) will now have $n_{\gamma_i}'=1$ (or one new state will have $n=2$). The final occupation numbers are thus $n_{\alpha_i}'$ and $n_{\gamma_i}'$, with (we give here only the case $n \ne m$ and $p \ne q$, the other cases being readily obtained)
\begin{eqnarray}
n_{\alpha_i}'=n_{\alpha_i} \quad \textrm{for} \quad i \ne n \quad \textrm{and} \quad i \ne m \nonumber \\
n_{\alpha_i}'=n_{\alpha_i}-1 \quad \textrm{for} \quad i=n \quad \textrm{or} \quad i=m          \nonumber \\
n_{\gamma_i}'=1 \quad \textrm{for} \quad i=p \quad \textrm{and} \quad i=q .
\label{nboccupfinal}
\end{eqnarray}
The number of initially non-populated untrapped states  $\gamma_i$ is very large, so that in both cases (BEC or thermal cloud), the terms with $p=q$ contibute in a negligible way. In addition, for the thermal case, as the number of populated trapped states is large, the terms with $n=m$ can also be safely neglected. For the thermal cloud, the dipole-dipole matrix element corresponding to transition from states $\alpha_n,\alpha_m$ to states $\gamma_p,\gamma_q$ thus reads:
\begin{eqnarray}
<Out|V|In> = \nonumber \\
N \frac{\exp (-\beta (E_{\alpha_n}+E_{\alpha_m})/2)}{ \sum_i \exp(-\beta E_{\alpha_i})} <\gamma_p \gamma_q |V_{dd} | \alpha_n \alpha_m>
\label{emBEC1}
\end{eqnarray}
and, for a condensate in the state $\alpha_0$,
\begin{eqnarray}
<Out|V|In> = \frac{N}{\sqrt{2}} <\gamma_p \gamma_q |V_{dd} | \alpha_0 \alpha_0>.
\label{emtherm1}
\end{eqnarray}
In both cases, the bras and kets of the matrix elements are now symmetrized two-atom wavefunctions.
In order to conveniently deal with the dipole-dipole interaction, we introduce the coupling of the spins, with total spin $\vec S_t=\vec S_1+\vec S_2$, and the separation of the center of mass and relative motions. Instead of the two-atom wavefunctions of equations (\ref{emBEC1}) and (\ref{emtherm1}), which are functions of the individual atom positions $\vec R_1$ and $\vec R_1$, we use atom pair wavefunctions of the new variables $\vec R_C=(\vec R_1+\vec R_2)/2$ and $\vec r=\vec R_1-\vec R_2$. These variables are indeed separable for the trapped motion in the harmonic trap as well as for the free motion. The two-body states, with wavefunctions $\phi_C$ and $\phi_{rel}$ describing the center-of-mass and relative motions, can be formally written in terms of the old ones

\begin{eqnarray}
|m_{S1},\alpha_1,m_{S2},\alpha_2> =\sum_{\phi_C,\phi_{rel}} |S_t, M_t, \phi_C ,\phi_{rel}>
\nonumber \\
<S_t, M_t,  \phi_C ,\phi_{rel}|m_{S1},\alpha_1,m_{S2},\alpha_2>, \nonumber \\
\textrm{with} \quad E_{\phi_C}+E_{\phi_{rel}}=E_{\alpha_1}+E_{\alpha_2}
\label{chgtbase}
\end{eqnarray}

The dipole-dipole matrix element is given by equations very similar to (\ref{emBEC1}) and (\ref{emtherm1}):
\begin{eqnarray}
<Out|V|In> = N \frac{\exp (-\beta (E_{\phi_C^{in}}+E_{\phi_{rel}^{in}})/2)}{ \sum_i \exp(-\beta E_{\alpha_i})}
\nonumber \\
<S_t',M_t',\phi_C^{out} \phi_{rel}^{out} |V_{dd} | S_t,M_t,\phi_C^{in} \phi_{rel}^{in}>,
\label{emtherm2}
\end{eqnarray}
for the thermal cloud, and
\begin{eqnarray}
<Out|V|In>=\frac{N}{\sqrt{2}} \nonumber \\
\times <S_t',M_t',\phi_C^{out} \phi_{rel}^{out}|V_{dd} |S_t,M_t, \phi_C^{in} \phi_{rel}^{in}>,
\label{emBEC2}
\end{eqnarray}
for the condensate. The ket in the latter equation is readily obtained, by variable change, from the two-atom wavefunction $|\alpha_0,\alpha_0>$.
In both cases, the dipole-dipole interaction does not concern the center of gravity motion and one has $\phi_C^{out}=\phi_C^{in}$. The decay rate, through dipolar relaxation, of the initial N-atom state is given by the Fermi Golden Rule (following first order perturbation theory)
\begin{eqnarray}
\Gamma=\frac{2\pi}{\hbar} \sum_{out} \left|<Out|V| In> \right|^2
\label{golden}
\end{eqnarray}
and is related to the atom loss by
\begin{eqnarray}
\Gamma=-\frac{1}{2}\frac{dN}{dt}
\label{atomloss}
\end{eqnarray}
In the BEC case one obtains
\begin{eqnarray}
\Gamma=\frac{2\pi}{\hbar}\frac{N^2}{2} \sum_{out} \left|<S_t',M_t',\phi_{rel}^{out}|V_{dd} |S_t,M_t,\phi_{rel}^{in}> \right|^2
\label{gamaBEC}
\end{eqnarray}
where $out$ in the sum stands for $S_t',M_t',\phi_{rel}^{out}$. Assuming that dipolar relaxation takes place in an otherwise non-interacting BEC, the relaxation rate parameter is finally given by
\begin{eqnarray}
\beta_{r}= \nonumber \\
\frac{4 \sqrt{2} \pi^{5/2}}{\hbar}a_{HO}^3 \sum_{out}
\left|<S_t',M_t',\phi_{rel}^{out}|V_{dd} |S_t,M_t,\phi_{rel}^{in}> \right|^2.
\label{betaBEC}
\end{eqnarray}
with $a_{HO}=\sqrt{\hbar / (m \omega)}$ the characteristic length of the trap. The initial and final wavefunctions $\phi_{rel}$ contain a spherical harmonic, with $l=m_l=0$, for the initial state and with $l=2$ and $m_l=M_t-M_t'$, for the final one, multiplied by a radial function.
As explained in the main part of the text, the radial part of the initial (resp. final) wavefunction behaves, at short distance, as a molecular vibrational function of the $S_t$ (resp. $S_t'$) molecular potential curve. At long distance, it behaves either as a gaussian wavefunction for the initial state, or as a free pair wavefunction, for the final one (eq. \ref{psiout}).

The case of an interacting BEC is harder to solve, as the relative and the center-of-mass motions of a pair of particles are not separable: the  wavefunction describing the relative motion of the atom pair is not independent of the relative directions of $\vec R_C$ and $\vec r$. For a BEC in the Thomas Fermi regime with radius $R_{TF}$,  the typical collision energy $\hbar^2/m R_{TF}^2$  is very small; we therefore assume an initial $l=m_l=0$ partial wave for any collision. We use an 'average' radial pair wavefunction which we calculate by evaluating  the radial density of probability $i.e.$ by the integration of the square of the two-atom Thomas-Fermi wavefunction over all the variables other than $|\vec r|$. The final Thomas-Fermi pair wavefunction is taken as the square root of this density of probability:
\begin{eqnarray}
F_{rel,TF}^{in}(R)= \nonumber \\ 
\frac{\sqrt{15}}{16 \sqrt{7}}~(2-R)^2 \sqrt{32+64R+24R^2+3R^3}.
\label{pairTF}
\end{eqnarray}
with $R=r/R_{TF}$. The relaxation rate parameter in the Thomas Fermi regime is given by
\begin{eqnarray}
\beta_{r}=\frac{28\pi^2}{15\hbar} R_{TF}^3 \sum_{out}
<S_t',M_t',\phi_{rel}^{out}|V_{dd} |S_t,M_t,\phi_{rel}^{in}> ^2.
\label{betaBECTF} 
\end{eqnarray}
where $\phi_{rel}^{in}$ is connected at short distance to a molecular wavefunction of $S_t$, and at long distance to the Thomas Fermi pair wavefunction (eq. \ref{pairTF}). We stress that the choice of the long distance radial wavefunction for the calculation of $\beta$ is of little importance, as $\beta$ involves an exact normalization to the atomic density. Hence, eq. (\ref{betaBEC}) and eq. (\ref{betaBECTF}) lead to the same numerical result.

In the case of a thermal cloud, both eq. (\ref{golden}) and (\ref{atomloss}) apply. However, the dipole-dipole interaction operator written in the second quantization form (eq. (\ref{sdquant})), when applied to the initial many-body state given by \ref{nboccup}, introduces another summation over the inital two-body states.  One has
\begin{eqnarray}
\Gamma=\frac{2\pi}{\hbar}N^2 \sum_{S_t',M_t',\phi_{rel}^{out},\phi_{rel}^{in}}  \frac{\exp(-\beta E_{in})}{\big(\sum_i \exp(-\beta E_{\alpha_i})\big )^2} \nonumber \\
\left|<S_t',M_t',\phi_{rel}^{out}|V_{dd} |S_t,M_t,\phi_{rel}^{in}> \right|^2
\label{gamatherm1}
\end{eqnarray}
where $E_{in}$ is the total energy of the initial state (including the motion of the center of gravity).
The center of mass motion and the relative one are both harmonic, with frequency $\omega/(2\pi)$. The initial pair states are characterized by quantum numbers $N_i,L_i,M_i$, with $E_{N_i,L_i}=(2N_i+L_i+3/2)~\hbar \omega$, for the global motion and $n_i,\ell_i,m_i$, with $E_{n_i,\ell_i}=(2n_i+\ell_i+3/2)~\hbar \omega$ for the relative one \cite{cdl}. It is important to notice that, because of symmetry properties, one has to restrict to even values of $\ell_i$, but that both even and odd  values of $L_i$ are allowed. The sum in the denominator of equation \ref{gamatherm1} still runs on atomic harmonic states. Provided that $\zeta=(\hbar \omega)/(k_BT)$ is small, one has
\begin{eqnarray}
D=\sum_i \exp(-\beta E_{\alpha_i})=\zeta ^{-3}
\label{sumD}
\end{eqnarray}
The equation (\ref{gamatherm1}) can be written as
\begin{eqnarray}
\Gamma=\frac{2\pi N^2}{\hbar D^2} \sum_{S_t',M_t',\ell',N_i,L_i,M_i,n_i,\ell_i,m_i}   \nonumber \\
\exp(-\zeta E_{N_i,L_i}) \exp(-\zeta E_{n_i,\ell_i}) \nonumber \\
<S_t',M_t',E',\ell',m'|V_{dd} |S_t,M_t,n_i,\ell_i,m_i> ^2
\label{gamatherm2}
\end{eqnarray}
where the energy $E'$ of the final state is fixed by the resonance condition and $m'=M_t-M_t'$. The sum over $N_i,L_i,M_i$ of $\exp(-\zeta E_{N_i,L_i})$ is equal to $D$ too. The atom loss for a gaussian atomic cloud is related to the relaxation rate by
\begin{eqnarray}
\frac{dN}{dt}=-\frac{1}{(4\pi)^{3/2}}~\beta_r~N^2~a_{HO}^{-3}~ \zeta^{3/2}
\label{loss2ratetherm}
\end{eqnarray}
The relaxation rate is thus given by
\begin{eqnarray}
\beta_{r}=\frac{(4\pi)^{5/2}}{\hbar}~a_{HO}^3 ~\zeta ^{3/2}\nonumber \\
\sum_{S_t',M_t',\ell',n_i,\ell_i,m_i} \exp(-\zeta (2n_i+\ell_i))\nonumber \\
\left|<S_t',M_t',E',\ell',m'|V_{dd} |S_t,M_t,n_i,\ell_i,m_i>\right| ^2.
\label{betatherm}
\end{eqnarray}
The matrix element in this equation can be split into angular and radial parts. The angular part yields the selection rules: triangular relations for the sets $\left(S_t,S_t',2\right)$, $\left(\ell_i,\ell',2\right)$, and $M_t'+m'=M_t+m_i$.
The radial integral is
\begin{eqnarray}
\label{rad-int2}
I(E')=\int_0^\infty \frac{F_{n_i,\ell_i}^{S_t} (r) F_{E',\ell'}^{S_t'} (r)}{r^3}~ r^2dr.
\end{eqnarray}
As explained in the main part of the text, the radial part of the initial (resp. final) wavefunction are connected at short distance, to a molecular vibrational function of the $S_t$ (resp. $S_t'$) molecular potential curve. 

We can now estimate the contribution of the different partial waves to dipolar relaxation. Figure \ref{partialwaves} shows the variation with $\Delta E$ of the square of the radial integrals for various initial and final quantum numbers. All integrals with $\ell_i \ne 0$ and $n_i \le 10$ vanish for a magnetic field larger than approximately 30 mG. For large enough energy gaps between initial and final states, the initial pair states with $\ell_i \ne 0$ do not contribute to relaxation because the effect of the rotational barrier on the final wave-function strongly limits the efficiency of the dipole-dipole interaction. As already mentioned in the main part of the text, although dipole-dipole interactions are long-range, only s-wave collisions contribute to dipolar relaxation at large enough magnetic fields.

We can also use eq. (\ref{betatherm}) to study the temperature dependence of dipolar relaxation. It is useful to introduce a radial wave function of a slightly different form (corresponding to a different normalization), $G(r)= r F(r)$, where $F(r)$ corresponds to the previously used radial wavefunction. Then, for $l=0$ initial states,  the matrix elements for dipolar relaxation are with a very good precision proportional to the derivative at $r=0$ of the long-range wavefunction corresponding to $G(r)$, $\sigma$, $i.e.$ its slope. This property arises from the fact that the amplitude of the wavefunction in the region where dipolar relaxation occurs is set by the long-distance wavefunction describing the motion in the trap (both are connected to describe both the molecular properties and the motion in the trap), which is  proportional to $\sigma$. This property stands both for harmonic oscillator and Thomas-Fermi wavefunctions.  The slope at the origin of the harmonic oscillator can easily be evaluated: $\sigma_{HO}(n_i) \sim \sqrt{\frac{8}{\pi}}~(n_i+1/2)^{1/4}~(a_{HO})^{-3/2}$ (for large $n_i$), and $\sigma_{HO}(0) = \left(\frac{2}{\pi}\right)^{1/4} a_{HO}^{-3/2}$. In case of a Thomas-Fermi pair wavefunction, $\sigma_{TF}=\sqrt{\frac{30}{7}}~R_{TF}^{-3/2}$

Introducing in equation (\ref{betatherm}) the proportionality to $(n_i+1/2)^{1/4}$ and to $(a_{HO})^{-3/2}$ of the matrix element, and summing over $n_i$, $\ell_i$ and $m_i$, we find that the relaxation rate of a thermal cloud is finally given by
\begin{eqnarray}
\beta_r=\frac{4\pi}{\hbar}\sum_{S_t',M_t'} 4\pi \nonumber \\
\frac{<S_t',M_t',E',2,M_t-M_t'|V_{dd} |S_t,M_t,0,0,0> ^2}{\sigma_{HO}(0)^2},
\label{formfintherm}
\end{eqnarray}
the fraction depending only on $S_t',M_t',E'$ and on the molecular parameters. It is now clear that the rate is independent of the temperature of the thermal cloud, provided $E'>>E$. We therefore understand why dipolar relaxation is temperature independent for large enough magnetic fields. This is in agreement with the results found in the framework of the first order Born approximation, neglecting all molecular potentials \cite{hensler}. This result remains valid even when molecular potentials are taken into account, and in particular in the vicinity of the dip in dipolar relaxation described in this paper. Such result is in agreement with our experimental observations shown in figure \ref{temperature}, which shows that the dipolar relaxation rate parameter is to within experimental uncertainty identical at $0.8 \mu$K and $4.5 \mu$K.

In figure \ref{temperature}, we also experimentally verify that the dipolar relaxation rate parameter is twice higher in a thermal gas than in a BEC. Our theoretical treatment also explains this fact. Indeed, the relaxation rate parameter of the non-interacting BEC (eq. \ref{betaBEC}) can be written as:
\begin{eqnarray}
\beta_r=\frac{2\pi}{\hbar}\sum_{S_t',M_t'} 4\pi \nonumber \\
\frac{<S_t',M_t',E',2,M_t-M_t'|V_{dd} |S_t,M_t,0,0,0> ^2}{\sigma_{HO}(0)^2};
\label{formfinBEC}
\end{eqnarray}
Comparing eq. (\ref{formfintherm}) and eq. (\ref{formfinBEC}), we deduce that the relaxation rate parameter for a thermal cloud is twice larger than for a condensate, as already known from, for example, \cite{verhaarstoof}. Such result holds in the dipolar relaxation dip (as shown in fig. \ref{temperature}). 

For low magnetic fields, the result would be different for two reasons. First, partial waves $\ell \ne 0$ contribute to dipolar relaxation in a thermal gas, and not in a BEC. Second, if the distance at which dipolar relaxation occurs is larger than the de Broglie wavelength, decoherence should be taken into account, and the coherent state chosen for the many-body input state (eq. (\ref{nboccup})) would not be a proper choice anymore.

The relaxation rate parameter of an interacting BEC in the Thomas Fermi regime can also be written in a similar way:
\begin{eqnarray}
\beta_r=\frac{2\pi}{\hbar}\sum_{S_t',M_t'} 4\pi \nonumber \\
\frac{<S_t',M_t',E',2,M_t-M_t'|V_{12} |S_t,M_t,\phi_{TF}> ^2}{\sigma_{TF}^2};
\label{formfinBECTF}
\end{eqnarray}
As the molecular wavefunction in the non interacting case and in the Thomas Fermi regime are proportional to respectively $\sigma_{HO}(0)$ and $\sigma_{TF}$ (the remaining of the wavefunction being identical in both cases, only set by $|S_t,M_t>$), we also confirm, by comparing eq. (\ref{formfinBEC}) and eq. (\ref{formfinBECTF}) that the result of the calculation of the dipolar relaxation rate parameter does not depend on the exact shape of the radial wavefunction, as stated above.

\section{Annexe II: Dipolar relaxation in 2D}

The Hamiltonian in the center of mass coordinate of two atoms reads
\begin{equation}
H=-\frac{\hbar^2}{m}\Delta+V_{\rm ext}(z)+V_{dd}(\vec r)+\frac{g_S \mu_B}{\hbar}\vec B\cdot\vec S 
\end{equation}
where $V_{\rm ext}(z)=(m/4)\omega_L^2z^2$, $\vec B$ is the magnetic field $\vec S=\vec S_1+ \vec S_2$ is the total spin and
$V_{dd}(\vec r)$ is the dipole-dipole interaction (see eq. (\ref{vdd})). $\omega_L / 2\pi$ corresponds to the vibrational frequency in a lattice site (see main part of the paper). 

The eigenstates of the Hamiltonian in absence of interaction read
\begin{equation}
\Psi_{\vec k, \mu}^n(\vec r)=e^{i\vec k\cdot \vec\rho}\phi_n(z)|\mu\rangle 
\end{equation}
with the energy
\begin{equation}
E_{\vec k, \mu}^n=\frac{\hbar^2k^2}{m}+\hbar\omega_L n -g\mu_BB \mu + \delta
\end{equation}
where $\vec{\rho} =(x,y)$, $\vec{k}=(k_x,k_y)$ is the momentum in the $xy-$ plane, $\phi_{n}(z)$
are the eigenstates of the harmonic oscillator in the $z$ direction and $|\mu\rangle $ are the eigenstates of the spin part of the Hamiltonian (defined in the main body of this paper), with $\mu=0,1,2$, and $\delta$ is a constant insuring that $-g\mu_BB \mu + \delta$ is the total magnetic energy of the atom pair.

The scattering cross section in two dimensions can be calculated in the first order Born approximation and reads
\begin{equation}
\label{scat}
\sigma^{n_i,n_f}_{\mu_i,\mu_f}(\vec k_i,\vec k_f)=\left(\frac{m}{2\hbar^2}\right)\frac{1}{2\pi k_i}\left|
\langle\psi_f|V_{dd}|\psi_i\rangle\right|^2|_{E_i=E_f}
\end{equation}
 $\vec k_i,n_i,\mu_i$ and $\vec k_f,n_f,\mu_f$ are the initial and final quantum numbers, and the energy must be conserved. The result is analytical.
The previous result is correct for distinguishable particles; in case of identical particles we have to replace
\begin{eqnarray}
\langle\psi_{\vec k_f, \mu_f}^{n_f}|V_{dd}|\psi_{\vec k_i, \mu_i}^{n_i}\rangle \rightarrow \\ \nonumber
\langle\psi_{\vec k_f, \mu_f}^{n_f}|V_{dd}|\psi_{\vec k_i, \mu_i}^{n_i}\rangle\pm\langle\psi_{\vec k_f, \mu_f}^{n_f}|V_{dd}|\psi_{-\vec k_i, \mu_i}^{n_i}\rangle
\end{eqnarray}
where the sign $+$ refers to bosons and $-$ to fermions.
In order to calculate the total cross cross-section we integrate and sum over all possible final states and we perform an average over the initial states with the same energy:
\begin{equation}
\sigma_{\mu_i,\mu_f}(k_i,k_f)=\frac{1}{2\pi}\sum_{n_f}\int\sigma^{n_i,n_f}_{\mu_i,\mu_f}(\vec k_i,\vec k_f)d\theta_fd\theta_i|_{E_i=E_f}
\end{equation}
where $\vec k_{i,f}=\left\|\vec k_{i,f} \right\| \left(\cos(\theta_{i,f}),\sin(\theta_{i,f})\right)$. 

The integral in eq.~(\ref{scat}) can be rewritten in the following form 
\begin{equation}
\langle\psi_f|V_{dd}|\psi_i\rangle=
\frac{1}{2\pi}\langle \mu_f|\tilde{V}_{dd}(\vec k_i-\vec k_f,q_z)|\mu_i\rangle \tilde{f}_{n_i,n_f}(-q_z)dq_z
\end{equation}
where $\tilde V_{dd}(\vec q)$ is the Fourier transform of the dipole-dipole potential $V_{dd}(\vec r)$
and $\tilde{f}_{n_i,n_f}(q_z)$ is the Fourier transform of $f_{n_i,n_f}(z)=(\phi_{n_f}(z))^*\phi_{n_i}(z)$. 

An important issue is the orientation of the magnetic field with respect to the 2D plane. Let us consider a magnetic field of the form
\begin{equation}
\vec B=B(\cos \theta \vec u_x+\sin \theta \vec u_z)
\end{equation}
In this case the matrix elements read
\begin{eqnarray}
&\langle 1|\tilde{V}_{dd}(\vec q)|0\rangle   \\
&=4\pi \frac{d^2}{q^2} \left(q_x\sin \theta-q_z\cos \theta+iq_y\right)\left(q_x\cos \theta+q_z\sin \theta\right) \nonumber
\end{eqnarray}
\begin{eqnarray}
&\langle 2|\tilde{V}_{dd}(\vec q)|0\rangle  \\
&=2\pi \frac{d^2}{q^2}(q_x\sin \theta-q_z\cos \theta+iq_y)^2 \nonumber
\end{eqnarray}
In the scattering process the total energy is conserved and all processes are not allowed; in particular the following condition must be fulfilled
\begin{equation}
\frac{\hbar^2k_f^2}{m}=\frac{\hbar^2k_i^2}{m}+\hbar\omega_L(n_i-n_f) - g_S \mu_B B (\mu_i-\mu_f)\geq0 ,
\end{equation}
which implies that there is magnetic field dependent upper bound for the final quantum number $n_f$. In our experiment the following conditions are fulfilled: $\hbar^2k_i^2/m<\hbar\omega_L$ and $n_i=0$. As a consequence if the magnetic field is not sufficiently intense the transition to higher state in the $z$-direction is forbidden. Since the difference of spin state can be $\mu_i-\mu_f=-1,-2$ the first transition occurs for $n_f=1$ and $\mu_i-\mu_f=-2$ at a critical magnetic field 
\begin{equation}
2 g_S \mu_B B_{cr 1}=\hbar\omega_L - \frac{\hbar^2k_i^2}{m}
\end{equation}
The second threshold appears for $n_f=2$ and $\mu_i-\mu_f=-2$ at a critical magnetic fields 
\begin{equation}
2 g_S \mu_B B_{cr 2}= 2\hbar\omega_L - \frac{\hbar^2k_i^2}{m}
\end{equation}
The next threshold appears for $n_f=1$ and $\mu_i-\mu_f=-1$ at a critical magnetic field
\begin{equation}
g_S \mu_B B_{cr 3}= \hbar\omega_L -  \frac{\hbar^2k_i^2}{m} 
\end{equation}
The second and the third thresholds coincide in the limit $\frac{\hbar k_i^2}{m\omega_L}\rightarrow0$. The first threshold corresponds to a pair of particle gaining a kinetic energy $\hbar\omega_L$ through dipolar relaxation in channel 2; after dipolar relaxation, each atom of this pair is then in a coherent superposition of the first two vibrationnal level of a lattice site.

An important aspect that strongly affects the inelastic scattering is the orientation of the magnetic field $\theta$. We can distinguish two limiting cases: the magnetic field orthogonal to the $xy-$plane, $\theta=\pi/2$ and the magnetic field lying in the $xy-$plane, $\theta= 0$. In fact different orientations of the magnetic field introduce different functional behaviors of $\sigma(B)$, and different selection rules since certain transitions are inhibited. The simplest example is the transition to $n_f=1$ with $\mu_i-\mu_f=-2$ for $\theta=\pi/2$; in this case the matrix element $\langle\psi_f|V_{dd}|\psi_i\rangle$ vanishes due to symmetry since the integration along the $z$ direction of an odd function vanishes. This is the reason why the first threshold disappears in fig \ref{reduceddimension} (b) (grey curve). Another important aspect is the limit of $B\rightarrow B_{cr}$: when $\theta=\pi/2$, the cross section of all channels opening above $B\rightarrow B_{cr}$ vanish, and this is the reason why the cross-section $\sigma(B)$ is then continuous. If $\theta \neq \pi/2$ the cross section of channels opening above $B\rightarrow B_{cr}$ do not generally vanish when $B\rightarrow B_{cr}$; as a consequence $\sigma(B)$ can present a discontinuous behavior. In practice the aspect of the first and second threshold is continuous, and the third is discontinuous. For the limit $B\rightarrow 0$ the behavior also differs as a function of $\theta$. For $\theta=\pi/2$, the dipolar relaxation rate parameter vanishes at low magnetic field and temperature $T=0$, whereas for $\theta= 0$, it is, to within a numerical factor,
\begin{equation}
\beta_{2D,T=0}^{\theta=0} \approx \frac{S^2 d^4 m}{\hbar^3} \frac{1}{a_L}.
\end{equation}


\begin{thebibliography}{99}



\bibitem{collapse} L. Santos, G. V. Shlyapnikov, P. Zoller, and M. Lewenstein, Phys. Rev. Lett. \textbf{85}, 1791 (2000), T. Lahaye, J. Metz, B. Fr$\ddot{o}$hlich, T. Koch, M. Meister, A. Griesmaier, T. Pfau, H. Saito, Y. Kawaguchi, and M. Ueda, Phys. Rev. Lett. \textbf{101}, 080401 (2008)

\bibitem{excitation} S. Yi and L. You, Phys. Rev. A \textbf{63}, 053607 (2001), D. H. J. O'Dell, S. Giovanazzi, and C. Eberlein, Phys. Rev. Lett. \textbf{92}, 250401 (2004), L. Santos, G. V. Shlyapnikov, and M. Lewenstein, Phys. Rev. Lett. \textbf{90}, 250403 (2003), S. De Palo, E. Orignac, R. Citro, and M. L. Chiofalo, Phys. Rev. B \textbf{77}, 212101 (2008) 

\bibitem{goral} K. G\'oral, L. Santos, and M. Lewenstein Phys. Rev. Lett. \textbf{88}, 170406 (2002)

\bibitem{baranov} M. A. Baranov, M. S. Mar'enko, V. S. Rychkov, and G. V. Shlyapnikov Phys. Rev. A \textbf{66}, 013606 (2002)

\bibitem{gaetan} A. Gaetan, Y. Miroshnychenko, T. Wilk, A. Chotia, M. Viteau, D. Comparat, P. Pillet, A. Browaeys, P. Grangier, Nature Physics \textbf{5}, 115 (2009)

\bibitem{jaksch} D. Jaksch, J. I. Cirac, P. Zoller, S. L. Rolston, R. C\^ot\'e, and M. D. Lukin Phys. Rev. Lett. \textbf{85}, 2208 (2000). See also D. DeMille, Phys. Rev. Lett. \textbf{88}, 067901 (2002)

\bibitem{ni} F. Lang et al., Phys. Rev. Lett. \textbf{101}, 133005 (2008), K. K. Ni et al., Science, \textbf{322}, 231 (2008).

\bibitem{jabez} J. J. McClelland and J. L. Hanssen Phys. Rev. Lett. \textbf{96}, 143005 (2006)

\bibitem{dy} Mingwu Lu, Seo Ho Youn, B. L. Lev, arXiv:0912.0050 (2009)

\bibitem{Griesmaier} A. Griesmaier et al., Phys. Rev. Lett. \textbf{94}, 160401 (2005)

\bibitem{beaufilsBEC} Q. Beaufils et al., Phys. Rev. A \textbf{77}, 061601 (2008)

\bibitem{fattoriflorence} M. Fattori, G. Roati, B. Deissler, C. D'Errico, M. Zaccanti, M. Jona-Lasinio, L. Santos, M. Inguscio, and G. Modugno, Phys. Rev. Lett. \textbf{101}, 190405 (2008)

\bibitem{Ho} Tin-Lun Ho, Phys. Rev. Lett., \textbf{81}, 742 (1998)

\bibitem{spindynamics} H. Schmaljohann et al., Phys. Rev. Lett.  \textbf{92}, 040402 (2004), M.-S. Chang et al., Phys. Rev. Lett. \textbf{92}, 140403 (2004), T. Kuwamoto et al., Phys. Rev. A \textbf{69}, 063604 (2004),  M. H. Wheeler et al., Phys. Rev. Lett. \textbf{93}, 170402 (2004), J. M. Higbie et al., Phys. Rev. Lett. \textbf{95},  050401 (2005).

\bibitem{dsk} M. Vengalatorre et al., Phys. Rev. Lett. \textbf{100}, 170403 (2008)

\bibitem{edh} Y. Kawaguchi, H. Saito, and M. Ueda, Phys. Rev. Lett. \textbf{96}, 080405 (2006), L. Santos and T. Pfau, Phys. Rev. Lett. \textbf{96}, 190404 (2006)

\bibitem{ueda1} Yuki Kawaguchi, Hiroki Saito, Masahito Ueda, Phys. Rev. Lett. \textbf{97}, 130404 (2006)

\bibitem{rydbergplasma} Wenhui Li, Paul J. Tanner, and T. F. Gallagher, Phys. Rev. Lett. \textbf{94}, 173001 (2005)

\bibitem{lithium} J. M. Gerton, C. A. Sackett, B. J. Frew, and R. G. Hulet Phys. Rev. A \textbf{59}, 1514 (1999)

\bibitem{helium} O. Sirjean, S. Seidelin, J. Viana Gomes, D. Boiron, C. I. Westbrook, A. Aspect, and G. V. Shlyapnikov Phys. Rev. Lett. \textbf{89}, 220406 (2002)

\bibitem{hensler} S. Hensler, J. Werner, A. Griesmaier, P. O. Schmidt, A. Gorlitz, T. Pfau, Appl. Phys B \textbf{77}, 765 (2002).

\bibitem{Lagendijk} A. Lagendijk, I. F. Silvera, and B. J. Verhaar Phys. Rev. B \textbf{33}, 626 (1986)

\bibitem{heliumth} G. V. Shlyapnikov, J. T. Walraven, U. M. Rahmanov, and M. W. Reynolds, Phys. Rev. Lett. 73, 3247 (1994), P. O. Fedichev, M. W. Reynolds, U. M. Rahmanov, and G. V. Shlyapnikov, Phys. Rev. A \textbf{53}, 1447 (1996)

\bibitem{fattori} M. Fattori, T. Koch, S. Goetz, A. Griesmaier, S. Hensler, J. Stuhler, and T. Pfau, Nature Physics \textbf{2}, 765 (2006)

\bibitem{landau} L. D. Landau and E. M. Lifshitz, Quantum Mechanics

\bibitem{who} F. H. Mies, E. Tiesinga, and P. S. Julienne, Phys. Rev. A \textbf{61}, 022721 (2000) 

\bibitem{werner} J. Werner, A. Griesmaier, S. Hensler, J. Stuhler, T. Pfau, A. Simoni, and E. Tiesinga, Phys. Rev. Lett. \textbf{94}, 183201 (2005)

\bibitem{Miller} J. D. Miller, R. A. Cline, and D. J. Heinzen, Phys. Rev. Lett. \textbf{71}, 2204 - 2207 (1993)

\bibitem{Cote} R. C\^ot\'e, A. Dalgarno, Y. Sun, and R. G. Hulet, Phys. Rev. Lett. \textbf{74}, 3581 - 3584 (1995)

\bibitem{verhaarstoof} H. T. C. Stoof, A. M. L. Janssen, J. M. V. A. Koelman, and B. J. Verhaar, Phys. Rev. A \textbf{39}, 3157 (1989)

\bibitem{d-wave-Fesh} Q. Beaufils Q., A. Crubellier, T. Zanon-Willette, B. Laburthe-Tolra, E. Mar\'echal, L. Vernac and O. Gorceix,
Phys. Rev. A, \textbf{79}, 032706 (2009)

\bibitem{crubellier} A. Crubellier, O. Dulieu, F. Masnou-Seeuws, M. Elbs, H. Kn$\ddot{o}$ckel, and E. Tiemann, Eur. Phys. J. D \textbf{6}, 211 (1999).

\bibitem{vanhaecke} N. Vanhaecke, Ch. Lisdat, B. T'Jampens, D. Comparat, A. Crubellier, P. Pillet, Eur. Phys. J. D \textbf{28}, 351 (2004).

\bibitem{dalibard} J. S$\ddot{o}$ding, D. Guery-Odelin, P. Desbiolles, F. Chevy, H. Inamori and J. Dalibard, Appl. Phys. B, \textbf{69}, 257 (1999).

\bibitem{beatriz} Beatriz Londo$\tilde{n}$o et al., in preparation.

\bibitem{hbt} T. Jeltes, J. M. McNamara, W. Hogervorst, W. Vassen, V. Krachmalnicoff, M. Schellekens, A. Perrin, H. Chang, D. Boiron, A. Aspect and C. I. Westbrook, Nature \textbf{445}, 402 (2007)

\bibitem{julienneconfinement} E. Tiesinga, C. J. Williams, F. H. Mies, and P. S. Julienne, Phys. Rev. A, \textbf{61}, 063416 (2000), E. L. Bolda, E. Tiesinga, and P. S. Julienne, Phys. Rev. A \textbf{66}, 013403 (2002).

\bibitem{petrovprl} D. S. Petrov, M. Holzmann, and G. V. Shlyapnikov, Phys. Rev. Lett. \textbf{84}, 2551 (2000)

\bibitem{petrovpra} D. S. Petrov and G. V. Shlyapnikov, Phys. Rev. A, \textbf{64}, 012706 (2001)

\bibitem{confinement} M. Olshanii, Phys. Rev. Lett. \textbf{81} 938 (1998), T. Bergeman, M. G. Moore, and M. Olshanii, Phys. Rev. Lett. \textbf{91}, 163201 (2003)

\bibitem{moritz} H. Moritz, T. St$\ddot{o}$ferle, K. G$\ddot{u}$nter, M. K$\ddot{o}$hl, and T. Esslinger Phys. Rev. Lett. \textbf{94}, 210401 (2005)

\bibitem{chuloss} V. Vuletic, A.J. Kerman, C. Chin, and S. Chu, Phys. Rev. Lett. \textbf{82}, 1406 (1999).

\bibitem{LiKrems} Z. Li and R. V. Krems, Phys. Rev. A, \textbf{79}, 050701(R) (2009)

\bibitem{bkt} N. Prokof'ev, O. Ruebenacker, and B. Svistunov, Phys. Rev. Lett. \textbf{87}, 270402 (2001)

\bibitem{bandmapping} A. Kastberg, W. D. Phillips, S. L. Rolston, R. J. C. Spreeuw, and P. S. Jessen, Phys. Rev. Lett. \textbf{74}, 1542 (1995)


\bibitem{luiten} O. J. Luiten, M. W. Reynolds, and J. T. M. Walraven, Phys. Rev. A \textbf{53}, 381 - 389 (1996).

\bibitem{verhaar} C. C. Agosta, I. F. Silvera, H. T. Stoof, and B. J. Verhaar, Phys. Rev. Lett. \textbf{62}, 2361 (1989), A. J. Moerdijk, B. J. Verhaar, and T. M. Nagtegaal Phys. Rev. A \textbf{53}, 4343 (1996)

\bibitem{beaufilslande} Q. Beaufils, T. Zanon, R. Chicireanu, B. Laburthe-Tolra, E. Mar\'echal, L. Vernac, J.-C. Keller, and O. Gorceix, Phys. Rev. A 78, 051603 (2008)

\bibitem{beaufilsrf} Q. Beaufils, T. Zanon, A. Crubellier, B. Laburthe-Tolra, E. Mar\'echal, L. Vernac et O. Gorceix,
Eur. Phys. J. D \textbf{56}, 99 (2010).

\bibitem{pilletrf} P. Pillet, R. Kachru, N. H. Tran, W. W. Smith, and  T. F. Gallagher, Phys. Rev. A \textbf{36}, 1132 (1987).

\bibitem{cohen}  C. Cohen-Tannoudji, J. Dupont-Roc and G. Grynberg, Atom Photon Interactions (Wiley, New York, 1992).

\bibitem{cdl} C. Cohen-Tannoudji, B. Diu and F. Lalo$\ddot{e}$, Quantum Mechanics, John Wiley, New York (1997).

\end{thebibliography}
\end{document}